\documentclass[twocolumn,trackchanges,apjl]{aastex631}

\usepackage{amsmath}
\usepackage{tabularx}
\usepackage{booktabs}
\let\tablenum\relax
\usepackage{siunitx}
\usepackage{multirow}
\usepackage{orcidlink}

\begin{document}

\title{Tracing the Host-Galaxy Environments of FRBs with a New Sample of Localized Sources Detected with the DSA-110}

\author[0009-0005-3565-4164]{Frank M. Verdi}
\affiliation{Cahill Center for Astronomy and Astrophysics, MC 249-17 California Institute of Technology, Pasadena CA 91125, USA.}

\author[0000-0002-7252-5485]{Vikram Ravi}
\affiliation{Cahill Center for Astronomy and Astrophysics, MC 249-17 California Institute of Technology, Pasadena CA 91125, USA.}
\affiliation{Owens Valley Radio Observatory, California Institute of Technology, Big Pine CA 93513, USA.}

\author[0000-0002-6573-7316]{Myles B. Sherman}
\author[0000-0001-9855-5781]{Jakob T. Faber}
\affiliation{Cahill Center for Astronomy and Astrophysics, MC 249-17 California Institute of Technology, Pasadena CA 91125, USA.}

\author[0000-0002-4941-5333]{Stella Koch Ocker}
\affiliation{Cahill Center for Astronomy and Astrophysics, MC 249-17 California Institute of Technology, Pasadena CA 91125, USA.}
\affiliation{Observatories of the Carnegie Institution for Science, Pasadena, CA 91101, USA.}

\author[0000-0002-4477-3625]{Kritti Sharma}
\affiliation{Cahill Center for Astronomy and Astrophysics, MC 249-17 California Institute of Technology, Pasadena CA 91125, USA.}

\author[0000-0002-4119-9963]{Casey J. Law}
\author[0000-0002-7083-4049]{Gregg Hallinan}
\affiliation{Cahill Center for Astronomy and Astrophysics, MC 249-17 California Institute of Technology, Pasadena CA 91125, USA.}
\affiliation{Owens Valley Radio Observatory, California Institute of Technology, Big Pine CA 93513, USA.}

\author[0000-0002-6823-2073]{Kaitlyn Shin}
\affiliation{Cahill Center for Astronomy and Astrophysics, MC 249-17 California Institute of Technology, Pasadena CA 91125, USA.}

\author[0000-0002-7587-6352]{Liam Connor}
\affiliation{Center for Astrophysics $\vert$ Harvard \& Smithsonian, Cambridge, MA 02138-1516, USA}

\collaboration{20}{(The Deep Synoptic Array team)}

\begin{abstract}

We characterize the source environments of a sample of fast radio bursts (FRBs) detected with the Deep Synoptic Array (DSA-110). We present new analyses of a sample of 24 FRBs which have been localized to host galaxies with spectroscopic redshifts. Four of these bursts have been confidently associated with galaxies near or beyond redshift 1, and several are published here for the first time. Combining these new bursts with previously analyzed DSA-110 sources, we use a sample of 43 DSA-110 FRBs with known redshifts to measure host-galaxy contributions to dispersion measure (DM) and Faraday rotation measure (RM). We find characteristically large and significantly correlated host-galaxy contributions to DM and RM, suggesting that most FRBs reside in overdense, magnetized regions within their host ISM. Additionally, we identify a subset of FRBs with detectable scattering, and show that their scattering timescales and inferred host DM values are broadly consistent with an empirical relation established for Galactic pulsars. Combined with two-screen modeling of some bursts, our results indicate that density fluctuations within the host-galaxy ISM are the dominant source of extragalactic scattering. Comparisons of FRB host-galaxy properties with pulsars and simulations of young and old progenitor populations support a scenario in which FRBs preferentially form and remain associated with H\,\textsc{ii} regions throughout their emission lifetime, favoring magnetars formed in core-collapse supernovae as the sources of FRBs. 

\end{abstract}

\section{Introduction} \label{sec:intro}

Fast radio bursts (FRBs; \citet{lorimer2007}) are luminous extragalactic radio transients characterized by large dispersion measures (DMs) in excess of Milky Way contributions and, in many cases, significant Faraday rotation measures (RMs). Although their formation channels, sources, and emission mechanisms remain uncertain, growing evidence suggests an association between FRBs and recent star formation. The discovery of similar radio bursts from the Galactic magnetar SGR 1935+2154 has offered further support for young magnetars as viable sources of FRB emission \citep{bochenek2020,chime2020}. With an increasing number of FRBs precisely localized to host galaxies, it has become possible to characterize their local environments as a population. Although FRBs have been found in a diverse range of galaxies and galaxy regions, recent work has found evidence for a preferential occurrence in massive star-forming galaxies, and regions of galaxies with high rates of star formation \citep{gordon2023,sharma2024,gordon2025}, suggesting a connection between the majority of FRB sources and young stellar progenitors. 

\begin{figure*}[!t]
    \centering
    \includegraphics[width=0.95\textwidth]{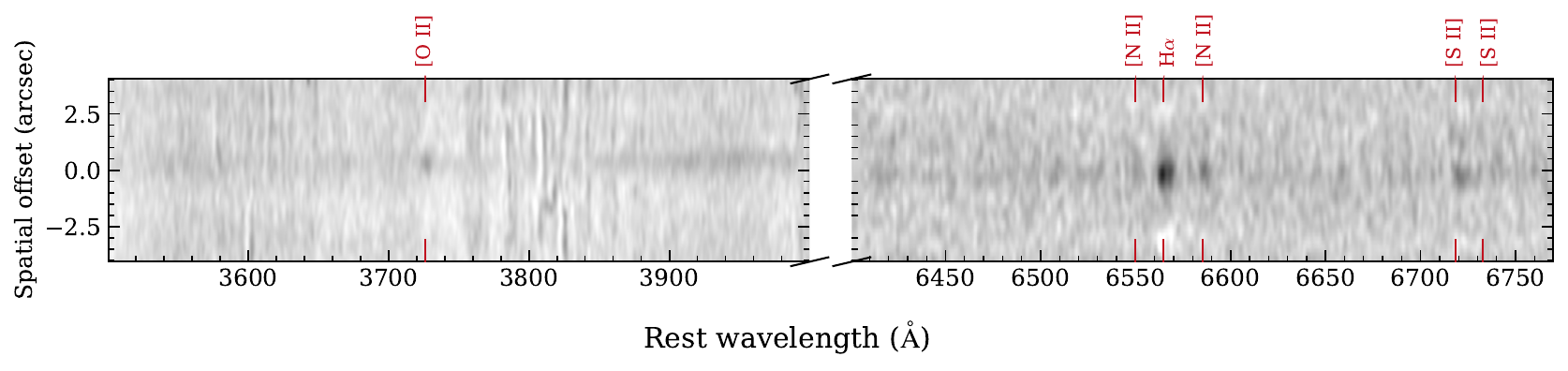}
    \caption{Two-dimensional optical/IR spectroscopy of the host galaxy of the DSA-110 discovered FRB\,20240104A. A redshift of $z=1.330$ was measured. The left panel displays the detection of the [OII] ($\lambda \lambda $ 3727, 3729) doublet with Keck-I/LRIS, and the right panel shows the detection of H${\rm \alpha}$ and the [SII] ($\lambda \lambda $ 6716, 6731) doublet with Keck-I/MOSFIRE. Details of the observations, and spectroscopy of two additional FRB hosts presented here for the first time, are given in the Appendix.}
    \label{fig:mike}
\end{figure*}

Despite their sources not being fully known, FRBs serve as powerful probes of the plasma along their lines of sight. FRB DMs quantify electron column densities between the sources and the observer, providing a direct measurement of baryon columns along the sightlines \citep{rickett1969,cordes1985}. These DMs are defined as
\begin{equation}
\text{DM}= \int_{0}^{L} n_e dl,
\label{eq:dm_def}
\end{equation}
where $n_e$ is the electron column density and $dl$ is the integral element along the line of sight. 

Similarly, FRB RMs probe the magnetic field components parallel to the propagation direction, weighted by the local electron density \citep{burn1966,brentjens2005}, and are given by
\begin{equation}
\mathrm{RM} = 0.812 \int_{0}^{L} n_e B_\parallel dl,
\label{eq:rm_def}
\end{equation}
where again $n_e$ is the electron column density and $B_{\parallel}$ is the magnetic field component along the line of sight.

We define RM$_{\text{host}}$ and DM$_{\text{host}}$ as the contributions to the total RM and DM originating from within the FRB host galaxy, measured in its rest frame. This includes the contribution from the host-galaxy interstellar medium (ISM) and any additional contributions from the immediate circumburst environment and the circumgalactic medium (CGM). The latter is expected to 
contribute little to DM$_{\text{host}}$ in typical FRB host galaxies \citep{connor2025}. If the source redshift is known and the intervening material is well-characterized, RM$_{\text{host}}$ and DM$_{\text{host}}$ can be constrained by subtracting foreground contributions. Prior work suggests that typical DM$_{\text{host}}$ values for FRBs may range from tens to a few hundred $\mathrm{pc\ cm^{-3}}$, seemingly consistent with modest column densities similar to the Milky Way ISM \citep{james2022,shin2023,orr2024,kovacs2024}. RM$_{\text{host}}$ has been found to span a remarkable range, from near zero to in excess of $10^5\ \mathrm{rad\ m^{-2}}$ for repeating FRBs, with some also demonstrating highly variable RMs \citep{plavin2022, anna-thomas2023}. Using a substantial catalog of well-localized FRBs and redshift measurements, we begin to characterize the underlying distribution of RM$_{\text{host}}$ and DM$_{\text{host}}$ across the population.

Combined measurements of RM and DM from the same medium also offer a direct probe of the line-of-sight integrated magnetic field \citep{masui2015,ravi2016}. For the magnetic field within the FRB host-galaxy, this relationship can be expressed as:
\begin{equation}
    \bar{B}_{\parallel, \text{host}} = \frac{\text{RM}_{\text{host}}}{0.81 \cdot \text{DM}_{\text{host}}} \,  \mathrm{\mu G}
    \label{eq:b_field}
\end{equation}
Therefore, using well-localized sources, it is possible to directly characterize extragalactic magnetic fields and thereby analyze the magnetism of FRB environments. These measurements can reveal signatures of dense and dynamic magneto-ionic environments (e.g., FRB 20121102) \citep{michilli2018,anna-thomas2023,mckinven2023A}, or probe the large-scale magnetic fields within the host-galaxy ISM \citep{sherman2023}, potentially constraining FRB source models and formation channels. 

Variations in plasma density also introduce scattering through multipath propagation, enabling FRBs to probe inhomogeneities in intervening media. This effect can manifest as an asymmetric, frequency-dependent temporal broadening of burst intensity profiles. For a Gaussian scattered image, this results in a characteristic exponential scattering tail in the pulse profile. Most FRBs exhibit scattering timescales $\tau(\nu)$ which greatly exceed those expected from Galactic contributions \citep{cordes2002}, suggesting that most pulse broadening has an extragalactic origin \citep{cordes2016,cordes2022,chawla2022}. Recent results favor FRB scattering dominated by scattering screens within the host-galaxy, arising either from inhomogeneities in the ISM or from immediate circumburst environments \citep{cordes2022,ocker2022,ocker2023}. Although scattering from intervening galaxy halos has been potentially observed \citep{faber2024}, and can result in extremely significant pulse broadening, these sightlines are uncommon. Evidence for scattering in the local environment has been observed in several FRBs \citep{abbott2025,ocker2023}, and growing evidence favors the ISM of the host galaxy as the dominant scattering medium \citep{masui2015,ocker2022,sammons2023}.

Motivated by the well-established empirical $\tau$--DM correlation observed for Galactic pulsars \citep{cordes2016,cordes2022}, previous studies have explored whether a similar correlation exists for FRB scattering \citep{ravi2019}. However, direct comparisons have been severely limited, as the $\tau$--DM relation for pulsars arises from Galactic DMs, and an appropriate comparison with FRB scattering therefore requires reliable constraints on DM$_{\text{host}}$ for a significant sample of FRBs. In this work, we leverage a large, well-localized sample of FRBs with robust redshift measurements to directly examine the $\tau$--DM$_{\text{host}}$ relationship.

In addition to pulse broadening, scattering in FRBs may also manifest as intensity variations in frequency with a characteristic scintillation bandwidth $\Delta \nu$. In practice, these bandwidths have been primarily resolved for scintillation caused by Milky Way scattering \citep[although see][]{ravi2016}, where the angular broadening of distant sources is sufficiently small that the scattered image remains unresolved by the Galactic screen \citep{masui2015,ocker2022,sammons2023,nimmo2025}. Constraining the distance to the dominant scattering medium for an individual FRB requires measurements of scintillation and/or pulse broadening on multiple, independent frequency scales. In this case, a two-screen scattering model can be applied under the assumption that these effects originate from two independent scattering screens, one Galactic and one extragalactic \citep{cordes2019, sammons2023,pradeep2025}. 

A comprehensive analysis of burst properties for a population of localized FRBs can therefore inform the electron density, magnetization, and inhomogeneity of source environments. In this paper, we analyze the burst properties of a sample of 24 FRBs detected with the DSA-110 which have been precisely localized to host-galaxies with spectroscopic redshifts. Several bursts have been confidently localized to high-redshift galaxies extending out to $z = 1.33$, including two of the most distant FRBs ever detected, and four new bursts near or beyond redshift 1. We publish four DSA-110 FRBs here for the first time, along with spectra of their host galaxies (see the Appendix and Figure~\ref{fig:mike}). We combine this new sample with previously published DSA-110 sources and other literature samples of well-localized FRBs to investigate FRB local environments. In §\ref{sec:burst_props}, we estimate burst properties for these new FRBs and isolate contributions from host galaxies. In §\ref{sec:rm_origin}, we use the RM$_{\text{host}}$--DM$_{\text{host}}$ correlation to investigate the origin of FRB rotation measures. In §\ref{sec:magnetic_fields}, we compare the magnetic field strengths traced by FRBs to the Milky Way ISM and to the elevated fields observed in H\,\textsc{ii} regions. In §\ref{sec:gal_sim}, we investigate the distribution and redshift dependence of DM$_{\text{host}}$, and compare these results to simulations of young and old FRB progenitor populations. In §\ref{sec:tau_dm}, we examine the $\tau$--DM$_{\text{host}}$ relationship for FRBs to identify the dominant source of extragalactic scattering. We discuss the implications of these results for FRB emission mechanisms, local environments, and progenitor types in §\ref{sec:discussion}, and conclude in §\ref{sec:conclusion}. Localizations and redshift estimates for previously published DSA-110 FRBs are presented in \citet{law2024}, \citet{sharma2024}, and \citet{connor2025}, while redshifts for new sources are obtained using the same methodology. Throughout this work, we adopt cosmological parameters from \citet{planck2016}. Data for all DSA-110 FRBs included in this paper can be accessed through the DSA-110 Archive\footnote{\url{https://code.deepsynoptic.org/dsa110-archive/}}.

\section{Inferring Burst Properties}
\label{sec:burst_props}

\subsection{DM$_{\text{host}}$ and RM$_{\text{host}}$}

We estimate DMs and RMs for the DSA-110 sample of FRBs using standard techniques. We assume that FRB dispersion arises primarily from electrons in a cold and sparse plasma. For FRBs with distinct sub-bursts, we measure RM using the sum of the on-burst spectra from all components. Following the methodology outlined in \citet{sherman2024}, we are able to confidently estimate new RMs for 28 FRBs in the DSA-110 sample we analyze herein. 

We assume that FRB DMs and RMs consist of the following contributions:
\begin{equation}
    \text{DM} = \text{DM}_{\text{MW}} + \text{DM}_{\text{MW,halo}} + \text{DM}_{\text{IGM}}(z) + \frac{\text{DM}_{\text{host}}}{(1+z)}
    \label{eq:dm_budget}
\end{equation}
\begin{equation}
    \text{RM} = \text{RM}_{\text{ion}} + \text{RM}_{\text{MW}} + \text{RM}_{\text{IGM}} + \frac{\text{RM}_{\text{host}}}{(1+z)^2}
    \label{eq:rm_budget}
\end{equation}
where ``ion'' indicates the contribution from Earth's ionosphere, ``MW'' refers to the Milky Way, and ``IGM'' includes contributions from ionized gas in the intergalactic medium, as well as intervening galaxy and cluster halos. We define ``host'' to include both the contribution from the host-galaxy ISM and any additional contribution from the immediate circumburst environment.

For each of the 43 well-localized FRB sources in the DSA-110 sample, we estimate DM$_{\text{host}}$ following the methods outlined by \citet{connor2023}. The probability density function (PDF) of DM$_{\text{host}}$ is derived by convolving the PDF of the observed DM with those inferred for each DM component. We assume that the observed DM has a Gaussian distribution with a standard deviation of $0.1 \, \mathrm{pc \, cm^{-3}}$ to account for both the small measurement uncertainty and potential systematic errors of DM measurement. Variations of this assumption within a reasonable range of measurement errors do not significantly affect our results. Following \citet{sherman2023}, we take DM$_{\text{MW}}$ to be a Gaussian distribution with a standard deviation of $30 \, \mathrm{pc \, cm^{-3}}$, centered about the Galactic contribution predicted by the NE2001 electron density model for the given line of sight \citep{cordes2002}. We further assume that DM$_{\text{MW,halo}}$ is described by a uniform distribution between $0$ and $40 \, \mathrm{pc \, cm^{-3}}$. DM$_{\text{IGM}}(z)$ takes the form described by \citet{connor2025} for the observed redshift, incorporating the expected contributions from intervening halos and the diffuse IGM. We treat the observed redshift as a constant parameter with negligible uncertainty ($<0.4\%$ \citet{sharma2023}). Following equation \eqref{eq:dm_budget}, the distribution of DM$_{\text{host}}$ in the observer frame is therefore given by the following convolution:
\begin{equation}
\begin{aligned}
    P\left( \frac{\text{DM}_{\text{host}}}{1+z} \right) &= P(\text{DM})  \circledast P(-\text{DM}_{\text{MW}}) \\
    &\quad  \circledast P(-\text{DM}_{\text{MW,halo}})  \circledast P(-\text{DM}_{\text{IGM}}(z))
    \label{eq:dm_conv}
\end{aligned}
\end{equation}

For bursts with measurable RMs, we estimate RM$_{\text{host}}$ using a similar method. Assuming a Gaussian distribution, the ionospheric contribution to the RM and its uncertainties are estimated for a given line of sight and arrival time using the RMExtract Python package, pulling from the NASA Archive of Space Geodesy Data \citep{mevius2018}. RM$_{\text{ion}}$ can fluctuate significantly depending on solar activity, but is typically of order $1 \, \mathrm{rad} \, \mathrm{m}^{-2}$, with uncertainties of order $0.1 \, \mathrm{rad} \, \mathrm{m}^{-2}$. As before, the PDF of the Galactic contribution, RM$_{\text{MW}}$, takes the form of a Gaussian, with the central value and uncertainty for each line of sight estimated using the \citet{hutschenreuter2022} map of the RM sky. These contributions are minimal, with typical values and uncertainties of order $10$–$20 \, \mathrm{rad} \, \mathrm{m}^{-2}$. The small magnetic fields of the IGM, generally no larger than several nano-Gauss \citep{hammond2012}, are not expected to contribute significantly to typical FRB RMs \citep{akahori2016,ravi2016,hackstein2019}. Therefore, following the initial conclusion of \citet{sherman2023} and the supporting results of this paper, which suggest that the extragalactic components of FRB RMs are dominated by the host-galaxy ISM, we neglect any RM contribution from IGM in our analysis.

 \begin{figure*}[t!]
  \centering
  \includegraphics[width=\textwidth]{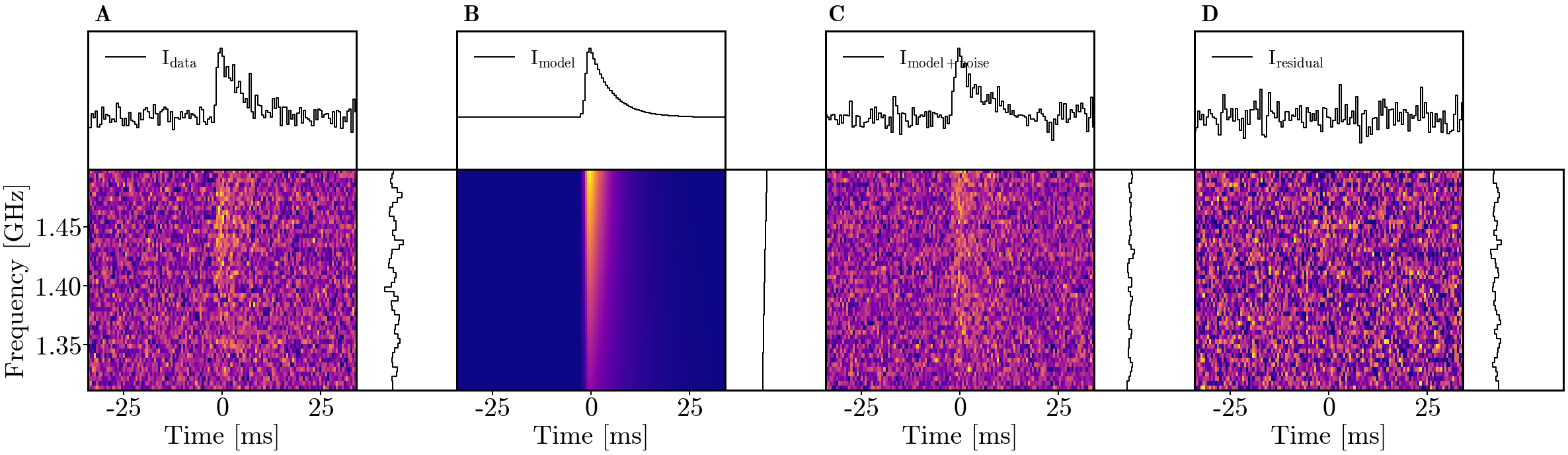}
  \caption{An example of the scattering model from Equation \eqref{eq:scattering_model} fitted to FRB 20231120A. \textbf{A} shows the dynamic spectrum with time and frequency profiles of the observed FRB; \textbf{B} shows the best-fit model \(S_\nu(t)\); \textbf{C} shows the same model with noise added at the level of the observed FRB; \textbf{D} shows the residual $S_\nu(t) - D_\nu(t)$ between the model and the data.}
  \label{fig:scattering_fit}
\end{figure*}

Following equation \eqref{eq:rm_budget}, we arrive at the following expression for the RM$_{\text{host}}$ distribution in the observer frame:
 \begin{equation}
    P\left( \frac{\text{RM}_{\text{host}}}{(1+z)^2} \right) = P(\text{RM}) \circledast P(-
    \text{RM}_{\text{ion}})  \circledast P(-
    \text{RM}_{\text{MW}}) 
    \label{eq:rm_conv}
\end{equation}
Using equation \eqref{eq:b_field}, these estimates of DM$_{\text{host}}$ and RM$_{\text{host}}$ are used to determine the probability distribution of $\bar{B}_{\parallel, \text{host}}$ for each FRB. The estimated host-galaxy contributions for all DSA-110 FRBs, along with the total RM and DM of each burst, is provided in Table~\ref{table:burst_props}. 

Intervening galaxies and galaxy clusters have also been shown to contribute significantly to the DM and RM of certain FRBs \citep{connor2023,lee2023,ramesh2023}, including one from this new DSA-110 sample \citep{faber2024}. We consider these potential contributions by querying the SIMBAD astronomical database \citep{wenger2000} for foreground galaxies near each sightline. From this search, we identify two FRBs (FRB 20231120A and FRB 20231220A) at offsets of $34\ \mathrm{kpc}$ and $78\ \mathrm{kpc}$ from intervening galaxies, which may intersect their halos. Previous work has shown that halos from individual galaxies at similar impact parameters contribute minimally to FRB DMs \citep{faber2024,khyrkin2024,connor2025,anna-thomas2025}, and upper limit estimates based on the Navarro–Frenk–White (NFW) profiles for these galaxies suggest maximum contributions of only a few $\mathrm{pc\ cm^{-3}}$. The expected contributions from halos along sightlines such as these are already captured by the probability distribution assumed for DM$_{\text{IGM}}(z)$ \citep{connor2025}. We therefore neglect the effect of these intervening halos when estimating DM$_{\text{host}}$ and RM$_{\text{host}}$. However, we cannot rule out the possibility that undetected galaxies or galaxy clusters may contribute more significantly to the observed DM and RM in some cases.

\subsection{Scattering Timescales}

We also estimate scattering timescales for the subset of bursts exhibiting clear pulse broadening with a one-sided exponential form. For all FRBs in this sample, we first search for temporal scattering by comparing two different pulse profile models fitted to each timeseries. Using a Monte Carlo fitting procedure, we first fit exponentially modified Gaussian functions $\mathbb{G}(t)$ to the frequency averaged timeseries of each burst, downsampled in time to increase S/N while preserving significant sub-structure. The form of this exponentially broadened pulse profile is given by 
\begin{equation}
\begin{aligned}
\mathbb{G}(t) = 
A \times 
\left[ \exp\!\left( -\frac{(t - \Delta t)^2}{2\sigma^2} \right) \right] \\
\circledast 
\left[ \mathbb{H}(t - \Delta t) \exp\!\left( -\frac{t - \Delta t}{\tau} \right) \right],
\label{eq:pulse_profile}
\end{aligned}
\end{equation}
where $t$ represents time, $\tau$ indicates the scattering timescale, $\mathbb{H}(t)$ is the Heaviside unit step function, $\Delta t$ is the time
shift, $\sigma$ is Gaussian standard deviation, and $\circledast$ indicates convolution. For bursts with multiple components, we modify this profile to include the sum of multiple gaussian functions, all convolved with the same one-sided exponential, under the assumption that the entire burst is scattered by a common intervening medium.

We then compare the Bayesian information criterion (BIC) of this model to that of an unmodified Gaussian model fit to the same timeseries. We consider a burst to have significant measurable scattering if the difference in BIC between the Gaussian model and the exponentially broadened Gaussian model is $\geq 6$. This procedure implicitly excludes low S/N bursts and bursts with pulse broadening timescales that do not significantly exceed the native DSA-110 time resolution of $32.768~\mu\mathrm{s}$. For bursts without measurable scattering timescales, we place an upper limit on $\tau$ using the standard deviation $\sigma$ of the best-fit unmodified Gaussian model, adopting the largest $\sigma$ among any burst components.

Using this method, we identify 15 FRBs within the complete DSA-110 sample that demonstrate statistically significant pulse broadening potentially attributable to scattering. For these bursts, we obtain more robust estimates of scattering timescales by fitting the following scattering model $S_\nu(t)$ to the dynamic spectra, accounting for frequency-dependent pulse broadening and changes in the intrinsic pulse profile across the band:
\begin{equation}
\begin{aligned}
S_\nu(t) &= \frac{c_\nu}{\sqrt{2 \pi \sigma_{\nu, \mathrm{DM}}^2}} \exp \left[\frac{-\left(t-\Delta t_{\nu, \mathrm{DM}}\right)^2}{\sigma_{\nu, \mathrm{DM}}^2}\right] \\
&\quad \circledast \mathbb{H}\left(t-t_0\right) \exp \left[\frac{t-t_0}{\tau_{1.4 \mathrm{GHz}} \nu^{-\alpha}}\right]
\label{eq:scattering_model}
\end{aligned}
\end{equation}
For each frequency $\nu$, $t$ represents time, $c_{\nu}$ is a coefficient that scales with the burst fluence as $\propto \nu^{-\gamma}$, where $\gamma$ is the spectral index. $\Delta t_{\nu,\ \mathrm{DM}}$ is the shift in time due to dispersive smearing after a start time $t_0$ referenced to $1530\ \mathrm{MHz}$, where $\Delta t_{\nu,\ \mathrm{DM}} = t_0 + t_{\nu, \mathrm{DM}}=4.15 \mathrm{~ms} \times \mathrm{\delta DM}\left(\nu / \mathrm{GHz}\right)^{-2}$. $\sigma_{\nu,\ \mathrm{DM}}$ is the Gaussian standard deviation that goes as $\propto \mathrm{DM}\left(\nu / \mathrm{GHz}\right)^{-3}$. $\tau$ represents the scattering timescale at 1.4\,GHz, which scales with frequency according to a scattering index $\alpha$, $\mathbb{H}(t)$ is the Heaviside unit step function, and $\circledast$ symbolizes a convolution. We fix $\alpha=4$, following expectations for a scattering medium with a Gaussian distribution of density fluctuations, or for a power-law density spectrum that is heavily suppressed below the diffractive scale. This choice is required for consistency with the exponential scattering tail assumed in the pulse broadening model \citep{geiger2025}. Reported scattering times have been scaled from 1.4 GHz to 1 GHz assuming $\alpha=4$ to enable comparison with other FRB and pulsar scattering catalogs. All values are given in the observing frame unless otherwise specified. 

We fit this frequency-dependent model to the dynamic spectrum of each scattered FRB using a Monte Carlo fitting procedure described in detail in \citet{faber2024}. We downsample the dynamic spectra from the native DSA-110 resolution in both time and frequency to increase S/N while preserving burst substructure prior to fitting this model. Figure~\ref{fig:scattering_fit} shows an example of $S_\nu(t)$ fitted to one burst in the sample, with fits for the remaining bursts presented in the Appendix. Table~\ref{table:burst_props} contains the values of $\tau$ at 1 GHz for each measurably scattered burst, as well as upper limits on $\tau$ for the remaining DSA-110 FRBs. 

\subsection{Scintillation Bandwidths}

For FRBs exhibiting pulse broadening, we also estimate the characteristic scintillation bandwidth $\Delta\nu$. The joint measurement of $\tau$ and $\Delta\nu$ is required to constrain the scattering geometry under the two-screen model, and we therefore do not consider other bursts exhibiting potential scintillation. The scintillation bandwidth is estimated as the half-width half-maximum (HWHM) of the intensity autocorrelation function (ACF) after accounting for a narrow spike at zero-lag from radiometer noise. We integrate the on-burst dynamic spectrum in time, preserving the original $0.0305\ \mathrm{MHz}$ frequency resolution, and calculate the on-burst ACF from this spectrum, excluding the zero-lag noise spike. The scintillation bandwidth is measured by fitting a Lorentzian to the central peak of the ACF using a least-squares fit. Resolvable scintillation bandwidths must exceed the $0.0305\ \mathrm{MHz}$ frequency resolution of the voltage measurements, and remain below the full $187\ \mathrm{MHz}$ bandwidth of the DSA-110. Using this method, we identify measurable scintillation bandwidths in 5 of the 15 bursts that exhibit pulse broadening, with the corresponding measurements shown in Figure~\ref{fig:appendix_scintillation}. No bursts from the sample show clear evidence of multiple scintillation scales. 

\subsection{Two-Screen Scattering Model}
For FRBs with measurable pulse broadening and scintillation, we constrain the distance between the source and a dominant local scattering medium assuming a two-screen model. We assume that $\tau$ originates from an extragalactic scattering screen, while $\Delta \nu$ results from a secondary scattering layer within the Milky Way \citep{ocker2022,cordes2022}. For this two-screen model, the observed pulse broadening and scintillation can be related to the scattering geometry by the following inequality:
\begin{equation}
L_x L_g \lesssim \frac{D_s^2}{2\pi \nu^2 (1+z)} \frac{\Delta \nu}{\tau}.
\end{equation}
where $\tau$ is the scattering timescale measured in the observing frame at 1 GHz, $\Delta \nu$ is the scintillation bandwidth, $D_s$ is the comoving distance between the observer and the source, and $\nu$ is the observing frequency in GHz. $L_g$ and $L_x$ are the distances from the observer to the Milky Way scattering layer and from the source to the extragalactic scattering layer, respectively, in kpc \citep{cordes2019,simard2021,ocker2022}. Combining measurements of $\tau$ and $\Delta \nu$ with the inferred $D_s$ from host-galaxy redshifts, we use this relationship to constrain $L_x L_g$ for five FRBs (Table~\ref{table:burst_props}). Adopting plausible values of $L_g$ based on the Galactic latitude of each FRB, expected to be of order $ 1\ \mathrm{kpc}$ \citep{ocker2020}, we estimate upper limits on $L_x$ ranging from several tens to several thousand kiloparsecs. However, our inability to resolve fine scintillation likely biases against more compact scattering geometries and prevents us from using this technique to distinguish between different sources of scattering within the host galaxy. 

\vspace{5mm}

\section{Properties of FRB Local Environments}
\label{sec:host_props}

\subsection{The Origin of FRB Rotation Measures}
\label{sec:rm_origin}

 \begin{figure}[b]
    \centering
    \includegraphics[width=0.475\textwidth]{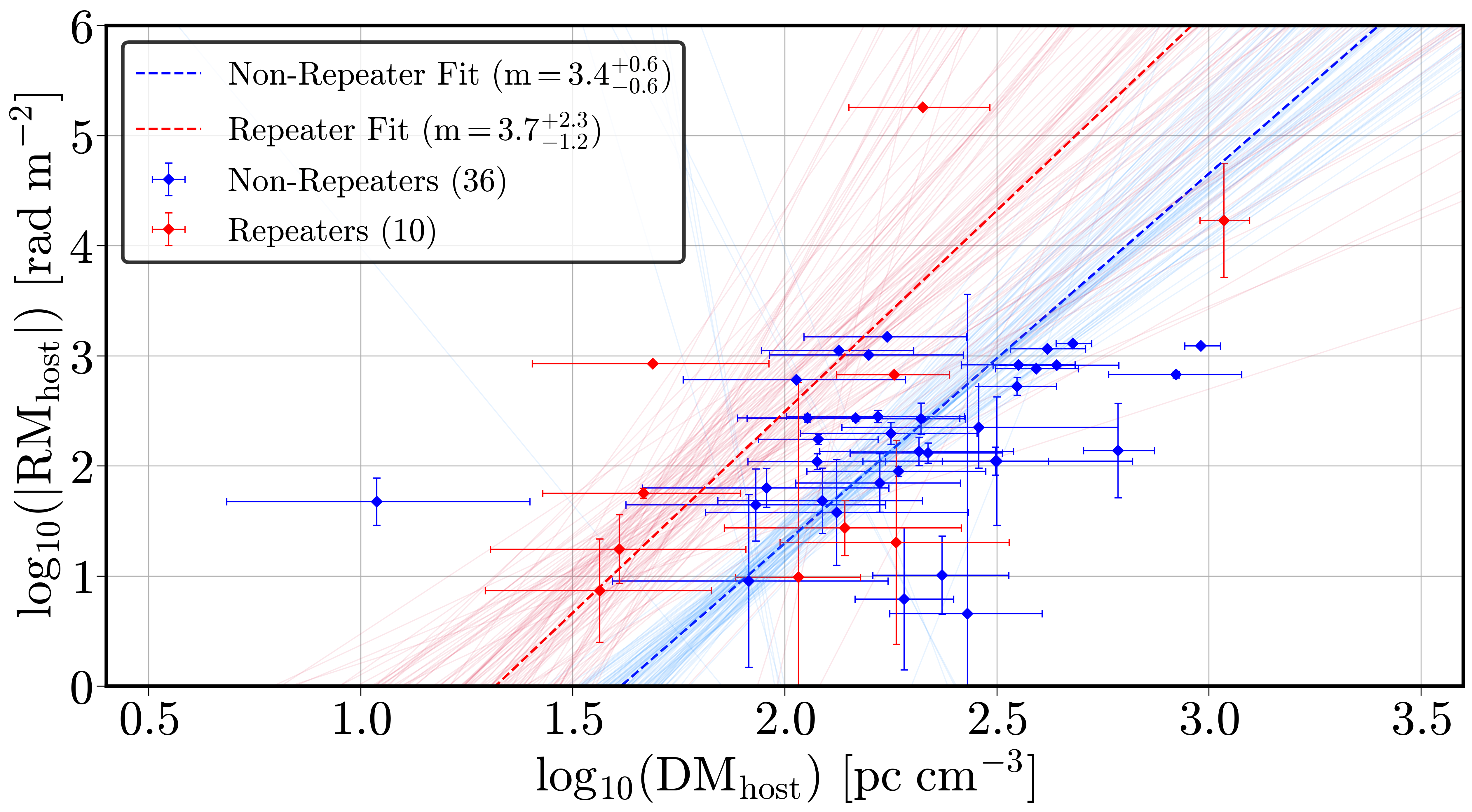}
    \caption{Estimates of \(\text{RM}_{\text{host}}\) and \(\text{DM}_{\text{host}}\) for as yet nonrepeating and repeating FRBs. Solid lines show the best fit from a Markov Chain Monte Carlo (MCMC) analysis for nonrepeaters and repeaters respectively, while faint lines show random draws from each simulation. The best-fit slopes with \(\pm1\sigma\) errors are displayed in the legend for both cases.}
    \label{fig:rm_dm}
\end{figure}

We first consider the relationship between RM and DM to investigate the origin of FRB rotation measures. If extragalactic FRB RMs are dominated by the host galaxy ISM, we expect a correlation between RM$_{\text{host}}$ and DM$_{\text{host}}$ \citep{sherman2023}, as both would be contributed primarily by the same medium. Conversely, a correlation can be masked if FRBs typically occupy dynamic local environments that generate significant RMs without correspondingly large DM contributions, such as the environment of FRB 20121102 \citep{michilli2018}. Tentative evidence for this correlation has been demonstrated by \citet{mannings2023} using a sample of nine FRBs, and by \citet{sherman2023} using a combined sample of 10 DSA-110 FRBs and 15 drawn from the literature. Here, we conduct a Spearman Rho test where the null hypothesis is that RM$_{\text{host}}$ and DM$_{\text{host}}$ are uncorrelated. We perform this test using a sample of 36 non-repeating DSA-110 FRBs with measured RMs and known redshifts, together with a sample of 10 repeating FRBs, including nine from the published literature one detected by the DSA-110. 

Figure~\ref{fig:rm_dm} shows that RM$_{\text{host}}$ and DM$_{\text{host}}$ are significantly correlated for nonrepeaters ($\rho= 0.44$, $p = 0.008$), repeaters ($\rho= 0.67$, $p = 0.03$), and for all FRBs ($\rho= 0.47$, $p = 0.0009$). When repeaters and nonrepeaters are considered as a single population, this suggests with 3-sigma confidence that the extragalactic RMs of FRBs are dominated by large-scale magnetic fields in the host-galaxy ISM, rather than highly magnetized and compact local environments. These results support, with increased confidence, the initial findings of \citet{mannings2023} and \citet{sherman2023}. 

\subsection{Magnetic Field Strengths}
\label{sec:magnetic_fields}

\begin{figure}[b]
    \centering
    \includegraphics[width=0.475\textwidth]{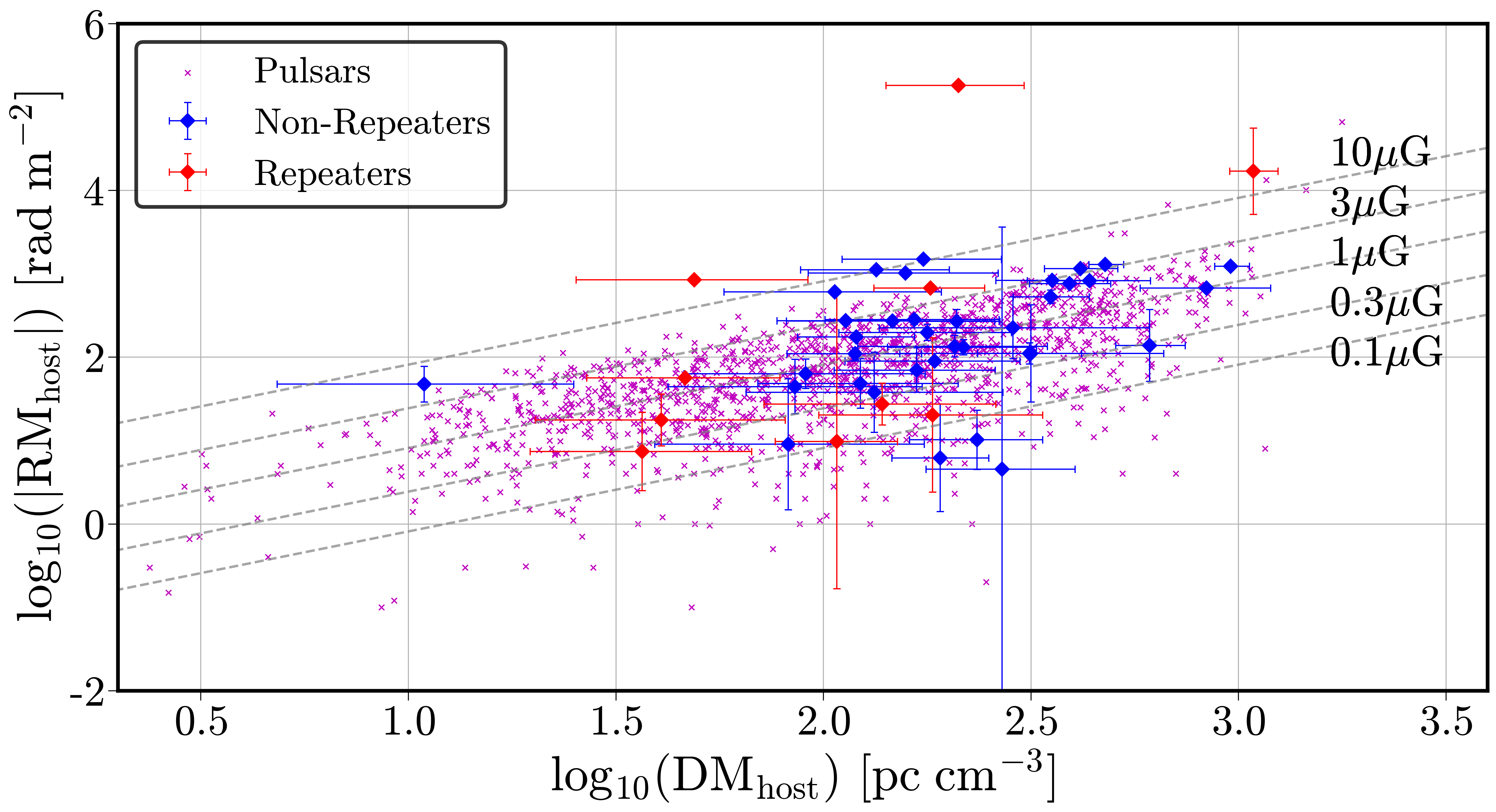}
    \caption{Comparison of FRB \(\text{RM}_{\text{host}}\) and \(\text{DM}_{\text{host}}\) estimates with the RMs and DMs of pulsars. Nonrepeating  and repeating FRBs are distinguished. Dotted gray lines of constant magnetic field strength, \(\text{B}_{\parallel, \text{host}}\) are shown.}
    \label{fig:rm_dm_pulsar}
\end{figure}

 Figure~\ref{fig:rm_dm_pulsar} shows RM$_{\text{host}}$ and DM$_{\text{host}}$ for FRBs compared to the total RMs and DMs of Galactic and Magellanic Cloud pulsars, including both canonical and millisecond pulsars. We compare the distributions of $\bar{B}_{\parallel, \text{host}}$ for FRBs and $\bar{B}_{\parallel}$ for pulsars using a Kolmogorov–Smirnov (KS) consistency test, with the null hypothesis that the two samples are drawn from the same distribution. We find significant differences between the magnetic fields probed by FRBs and Galactic pulsars ($D = 0.2$, $p = 0.01$ for all FRBs, $D = 0.2$, $p = 0.07$ for nonrepeaters, and $D = 0.4$, $p = 0.07$ for repeaters). The $\bar{B}_{\parallel}$ distribution inferred for DSA-110 FRBs is characteristically $\sim1–2 \mu\text{G}$ larger than for Galactic pulsars, with non-repeating FRBs tracing $\bar{B}_{\parallel} > 6 \mu\text{G}$ at more than 20 times the rate of pulsars (11\% of nonrepeaters, 30\% of repeaters, and 0.5\% of pulsars.) Having established that $\bar{B}_{\parallel}$ likely traces magnetic fields of the large-scale host-galaxy ISM, we conclude that FRBs may tend to occupy galaxies or local environments with greater ISM magnetic-field strengths than the Milky Way, and that the most highly magnetized environments are populated at a significantly greater rate by FRBs than by pulsars.

 Cosmological simulations of FRB populations distributed within a smooth ISM predict that such sources would rarely probe $\bar{B}_{\parallel} \gtrsim 3\ \mu\text{G}$ \citep{kovacs2024}. Instead, the distribution of $\bar{B}_{\parallel}$ inferred from this sample of FRBs may be consistent with the amplified magnetic fields traced in regions of increased star-formation, such as H\,\textsc{ii} regions. RM measurements of Galactic and extragalactic background sources have been used to estimate $\bar{B}_{\parallel}$ for a sample of foreground Galactic H\,\textsc{ii} regions \citep{heiles1980,heiles1981,mitra2003,harveysmith2011}. We note that magnetic field measurements of H\,\textsc{ii} regions are limited in number and may be affected by selection biases. High-resolution cosmological simulations such as FIRE \citep{hopkins2018} that resolve individual star-forming regions will be necessary for obtaining more robust estimates of magnetic field strengths in H\,\textsc{ii} regions, and for exploring this comparison in further detail.  
 
\begin{figure}[t]
    \centering
    \includegraphics[width=0.475\textwidth]{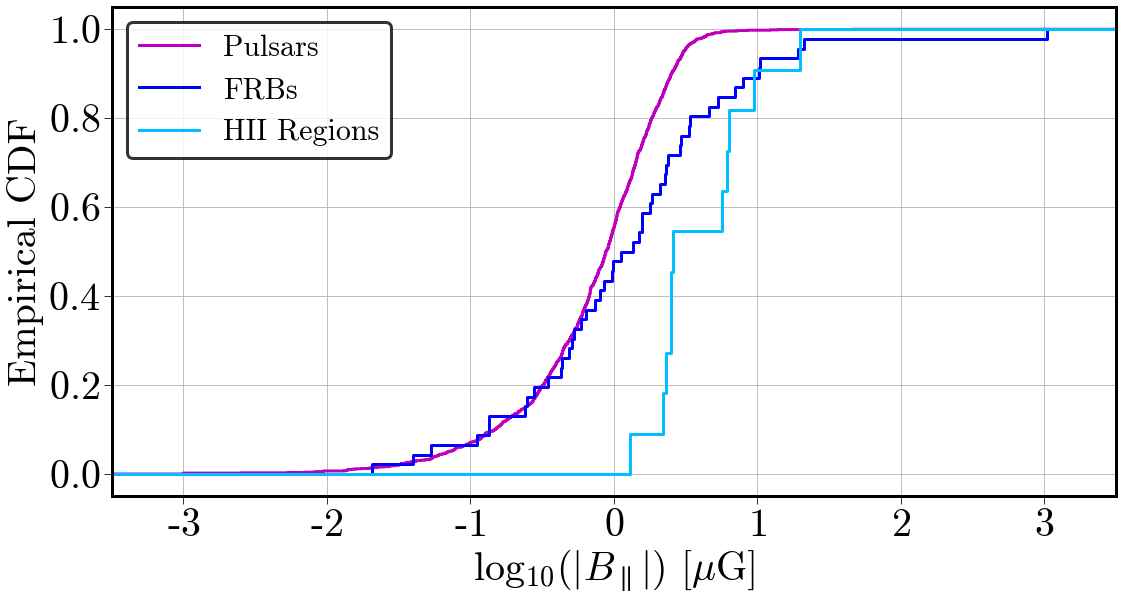}
    \caption{Distributions of line-of-sight magnetic field strengths $\bar{B}_{\parallel}$ for FRB host galaxies, Galactic pulsars, and Galactic H\,\textsc{ii} regions, shown as empirical cumulative distribution functions (CDFs). Nonrepeating and repeating FRBs are treated as a single population. FRBs trace large-scale magnetic fields that are often stronger than those in the diffuse Milky Way ISM probed by pulsars, and comparable to field strengths estimated for H\,\textsc{ii} regions.}
    \label{fig:b_distribution}
\end{figure}

 In Figure~\ref{fig:b_distribution}, we compare $\bar{B}_{\parallel}$ for FRB host-galaxies with pulsars tracing the Milky Way ISM and for contributions from Galactic H\,\textsc{ii} regions. We find that a considerable fraction of FRBs trace line-of-sight magnetic field strengths exceeding those of Galactic pulsars, but comparable to those inferred for H\,\textsc{ii} regions. While the FRB distribution is not statistically consistent with either comparison population ($p = 0.03$ for Galactic pulsars, $p = 0.004$ for H\,\textsc{ii} regions, using a KS consistency test), it spans a broad range of magnetic environments, from typical diffuse ISM conditions to the enhanced fields associated with H\,\textsc{ii} regions. The observed distribution may naturally arise if FRBs are commonly embedded in H\,\textsc{ii} regions but often sample only a portion of their full magnetic field, due to variations in sightline geometry and field reversals within the host galaxy. In the following section, we investigate the redshift evolution of these host-galaxy contributions to further assess whether this sample of FRBs is consistent with a population tracing regions of elevated star formation. 

While the strong RM$_{\text{host}}$--DM$_{\text{host}}$ correlation in the sample suggests that $\bar{B}_{\parallel}$ traces the magnetic field of the local ISM for most FRBs, a subset of bursts are known to receive significant RM contributions from highly magnetized and dynamic circumburst environments. This type of circumburst environment has been previously demonstrated for two repeating FRBs in the literature sample with large ($|\mathrm{RM}_{\text{host}}| > 10^4$) and significantly varying RMs \citep{michilli2018,plavin2022, anna-thomas2023}. Large-scale magnetic fields in the ISM are expected to saturate at only several $\mu\text{G}$ \citep{su2018,rodrigues2019}, and these two FRBs probe $\bar{B}_{\parallel}$ considerably exceeding this theoretical limit. Thus far, no comparable outliers have been observed among nonrepeating FRBs.

\subsection{Comparison with Simulated Galaxy Populations}
\label{sec:gal_sim}

\begin{figure}[t]
  \centering
  \includegraphics[width=0.475\textwidth]{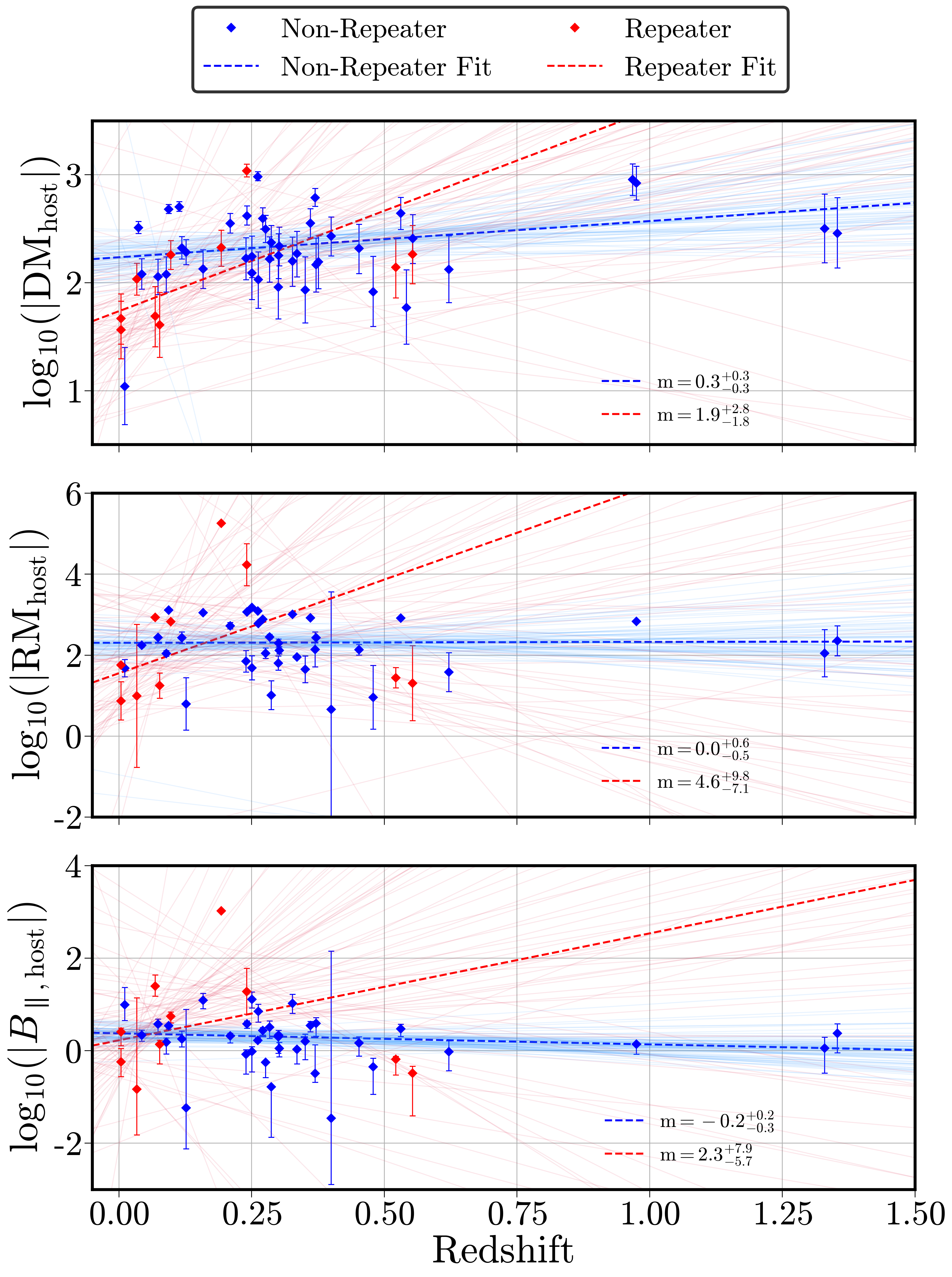}
  \caption{\(\text{DM}_{\text{host}}\), \(\text{RM}_{\text{host}}\), and \(\text{B}_{\parallel, \text{host}}\) for FRBs at different redshifts. Dashed lines show the best fit from an MCMC analysis and random draws from each simulation are shown by faint lines. Note that values are scaled to the host galaxy rest frame, and that redshift error bars are omitted.}
  \label{fig:redshift_trends}
\end{figure}

We now investigate and interpret the distribution of host-galaxy properties and their correlation with redshift. This analysis is significantly bolstered by the inclusion of several high-redshift FRBs in the updated DSA-110 sample, which extends to $z = 1.33$ and includes four new bursts near or beyond redshift 1. DM$_{\text{host}}$, RM$_{\text{host}}$, and $\bar{B}_{\parallel, \text{host}}$ have been corrected for redshift suppression, and therefore any observed correlation reflects a physical change in host-galaxy properties at increasing redshift. To scale these quantities to the rest frame, we assume that DM and $\bar{B}_{\parallel}$ scale with $(1+z)^{-1}$, while RM scales with $(1+z)^{-2}$. We find no significant correlation with redshift for DM$_{\text{host}}$, RM$_{\text{host}}$, or $\bar{B}_{\parallel}$, as shown in Figure~\ref{fig:redshift_trends}. 

\begin{figure}[t]
    \centering
    \includegraphics[width=0.47\textwidth]{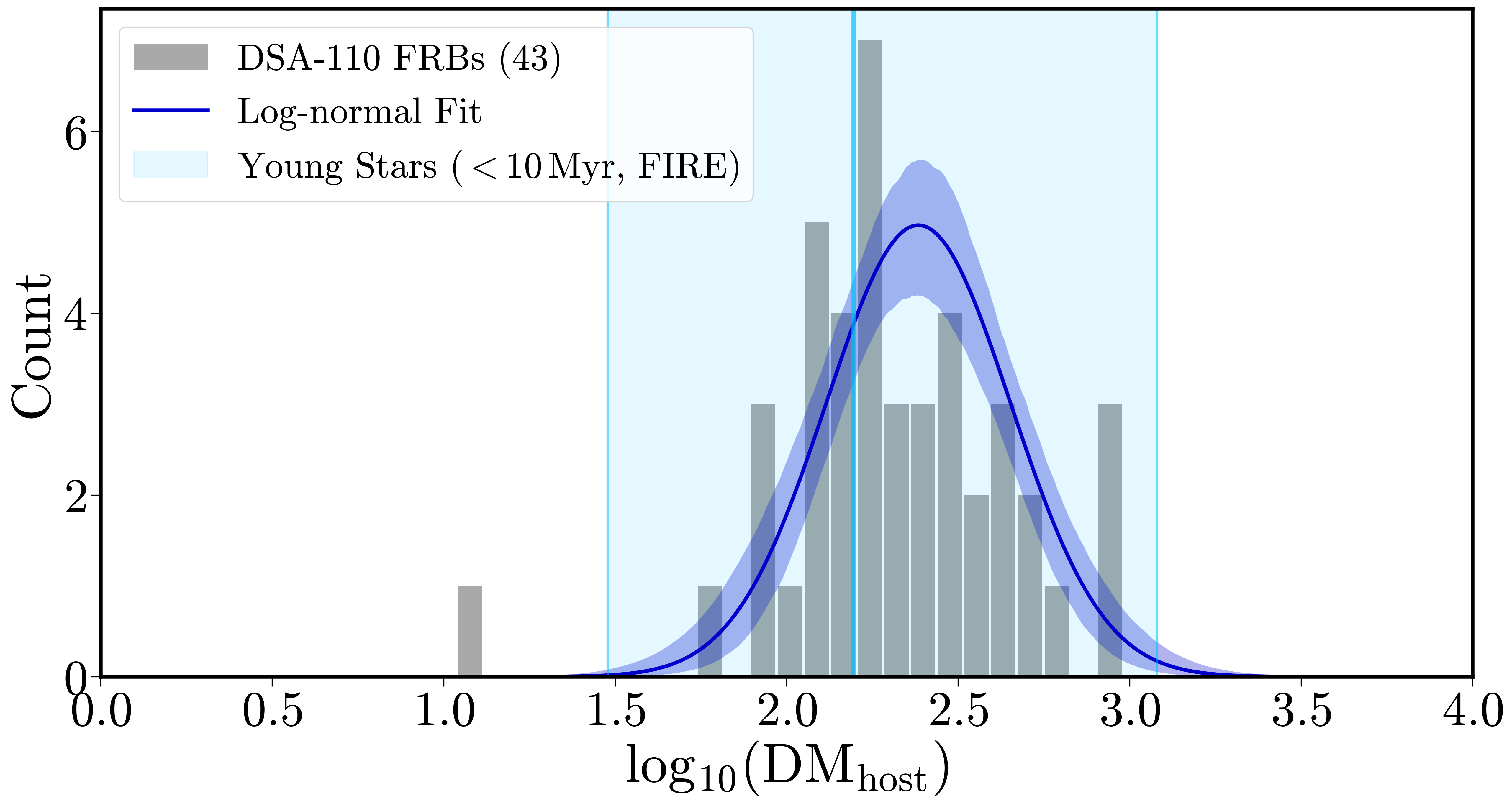}
    \caption{Distribution of log-scaled DM\(_{\text{host}}\) for the DSA-110 sample, fit to a lognormal distribution. The shaded dark blue indicates the $68\%$ credible interval of the fit. Overlaid in light blue is the range of values predicted for FRBs with young stellar progenitors (\(<10\) Myr; 90\%) at $z=0$ using the FIRE-2 simulation \citep{orr2024} of orientation averaged galaxies, with the corresponding median marked by a light blue vertical line.}
    \label{fig:dm_distribution}
\end{figure}

To model the observed distribution, we fit a probability density function (PDF) to the sample of DM$_{\text{host}}$ measurements from DSA-110 FRBs. We assume that the PDF of DM$_{\text{host}}$ for the FRB population can be described by a lognormal distribution
\begin{equation}
f_{\mathrm{DMHost}}(x) = \frac{1}{\sqrt{2\pi}\sigma x} \exp\left[-\frac{(\ln x - \mu)^2}{2\sigma^2}\right],
\label{eq:log_norm}
\end{equation}
where $x = \text{DM}_{\mathrm{host}}$, and $\mu$ and $\sigma$ are the mean and standard deviation of $\ln({\mathrm{DM}_{\text{host}})}$, respectively. This form is motivated by the lognormal PDFs observed in the column density of both neutral and ionized gas in the local Milky Way ISM \citep{imara2016,bialy2019}. We fit this model to the data using MCMC sampling, taking the likelihood of each FRB to be the uncertainty distribution of its DM$_{\text{host}}$ convolved with the PDF for the population. The resulting posterior medians are given by $\mu = 5.49^{+0.12}_{-0.12}$ and $\sigma = 0.62^{+0.11}_{-0.09}$. A histogram of DM$_{\text{host}}$ for DSA-110 FRBs is shown in Figure~\ref{fig:dm_distribution} alongside this best-fit probability distribution. 

To interpret this distribution of DM$_{\text{host}}$, we compare these results to the predicted distributions from different cosmological galaxy-formation simulations. We first consider the host-galaxy contributions to DM estimated by \citet{kovacs2024} using the TNG50 simulation from the IllustrisTNG project. \citet{kovacs2024} considered a total of 16,500 simulated galaxies in the stellar mass range of $\log{(M_{\ast}/M_{\odot}})=9$--$12$ and with non-zero star formation rates. For each of these simulated galaxies considered, 1000 FRB positions were drawn from random gas cells within a radius enclosing 99\% of the galaxy's star formation. From this population of simulated FRBs, probability density functions of DM$_{\text{host}}$ and RM$_{\text{host}}$ were constructed for galaxies at different redshifts. They report that DM$_{\text{host}}$ increases significantly with redshift, owing primarily to an increase in ISM electron densities caused by the higher star formation rate (SFR) of older galaxies. Inferred values of DM$_{\text{host}}$ for the FRB population are significantly greater than predicted by TNG50, often by hundreds of pc cm$^{-3}$. Furthermore, we observe considerably more variance in DM$_{\text{host}}$ and RM$_{\text{host}}$, and do not find evidence for the expected DM--$z$ correlation.

\begin{figure}[t]
    \centering
    \includegraphics[width=0.47\textwidth]{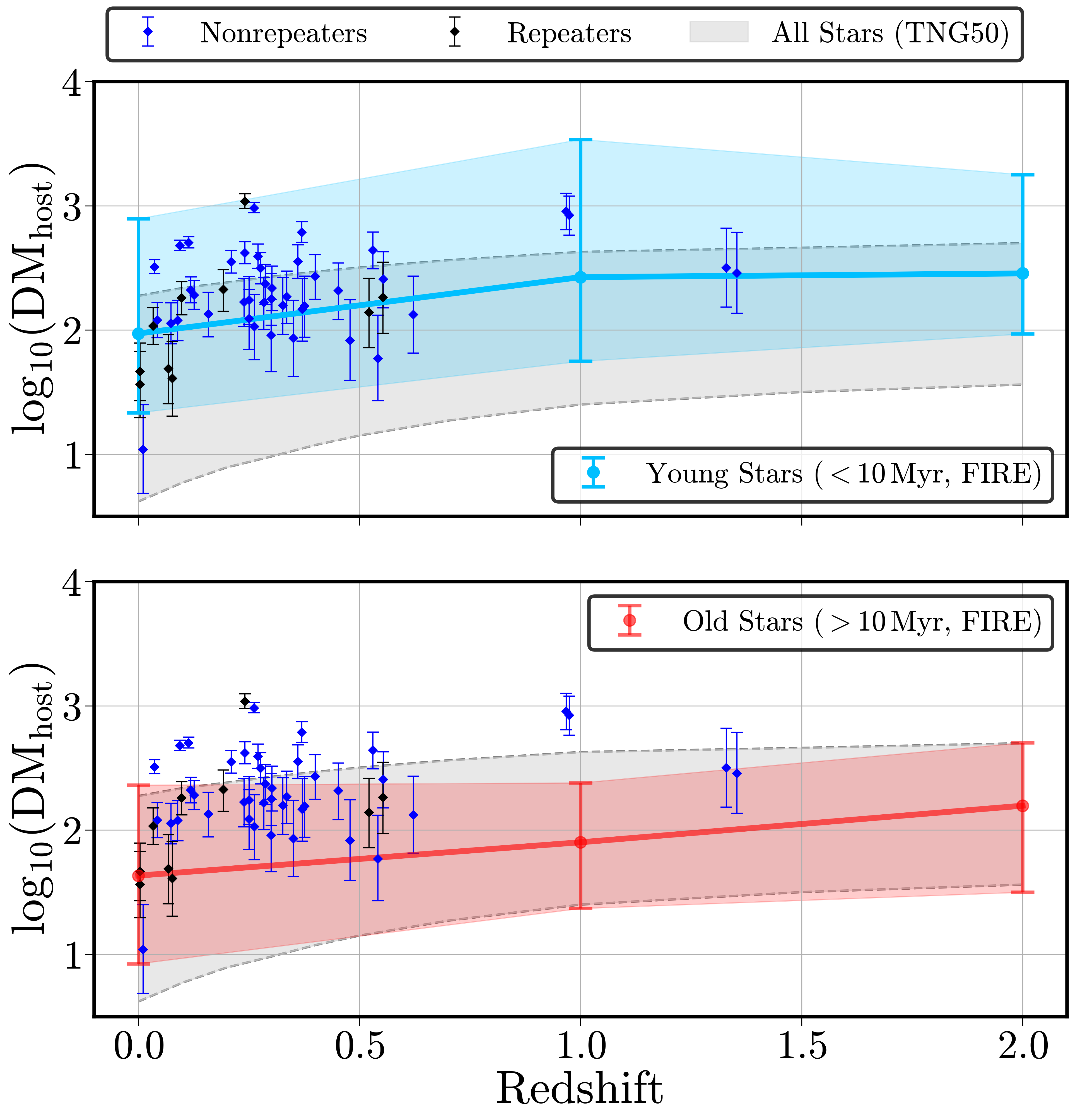}
    \caption{Redshift evolution of log-scaled DM\(_{\text{host}}\) for the DSA-110 and literature FRB sample compared to predictions from FIRE-2 and TNG50 simulations \citep{orr2024,kovacs2024}. The top panel shows DM\(_{\text{host}}\) estimates plotted against redshift, overlaid with the distribution simulated for FRBs with young stellar progenitors (\(<10\) Myr; blue, 90\%) in face-on galaxies at three redshift snapshots ($z=0,1,2$). The bottom panel shows the same data compared to simulations of older stellar populations from FIRE (\(>10\) Myr; red, 90\%) and TNG50 (gray, 90\%). The elevated and more varied distribution of DM\(_{\text{host}}\) aligns more closely with the young star model, suggesting a potential link between FRBs and regions with high star formation rates.}
    \label{fig:fire_comp}
\end{figure}

We then compare the DSA-110 sample to the DM$_{\text{host}}$ distributions modeled by \citet{orr2024} using the Feedback in Realistic Environments (FIRE-2) simulation \citep{hopkins2018}. This study considers only 6 simulated MW-mass galaxies (and their time-evolution snapshots at $z = 0,1,2$), but has the ability to resolve individual H\,\textsc{ii} regions at parsec scales, rather than treating the ISM as a smooth equation-of-state. Two different FRB progenitors are considered: a young population ($<10\,\mathrm{Myr}$) embedded primarily within H\,\textsc{ii} regions and an older population ($>10\,\mathrm{Myr}$) outside of H\,\textsc{ii} regions, differentiated by how recently each star was formed in the simulation. The young population exhibits a significantly higher median DM and broader DM distribution than the old population, driven by young progenitors remaining clustered within overdense H\,\textsc{ii} regions of the ISM. Comparing the results of this simulation to DSA-110 FRBs, we find that our distribution of DM$_{\text{host}}$ closely matches the predicted distribution for young progenitors embedded in regions of active star formation (see also Law et al., in prep.). Comparisons of the DSA-110 sample to these simulated distributions are presented in Table~\ref{tab:dm_fire2} and Figure~\ref{fig:fire_comp}. 

\begin{deluxetable*}{lccccc}[t]
\label{dm_table}
\tabletypesize{\footnotesize}
\tablecaption{DM$_{\text{host}}$ Percentiles from the Lognormal Fit to DSA-110 FRBs and from FIRE-2 Galaxy Simulations \label{tab:dm_fire2}}
\tablehead{
\colhead{} & \colhead{\textbf{Population}} & \colhead{\textbf{5\%}} & \colhead{\textbf{50\%}} & \colhead{\textbf{Mean}} & \colhead{\textbf{95\%}}
}
\startdata
\noalign{\vskip 4pt}
& {\textbf{DSA-110 FRBs}} & $87.4^{+20.3}_{-18.0}$ & $242.2^{+29.4}_{-27.5}$ & $293.6^{+38.9}_{-31.5}$ & $666.0^{+140.5}_{-99.9}$ \\
\noalign{\vskip 4pt}
\hline
\multirow{4}{*}{\textbf{FIRE-2 ($z=0$)}} 
        & \centering Young (face-on)          & 21.6 & 93.6 & 210  & 787 \\
        & \centering Old (face-on)            & 8.4  & 43.0 & 73.8 & 230 \\
        & \centering Young (orientation avg.) & 30.1 & 157  & 328  & 1203 \\
        & \centering Old (orientation avg.)   & 11.5 & 84.4 & 149  & 488 \\  
\noalign{\vskip 4pt}
\hline
\noalign{\vskip 4pt}
\multirow{2}{*}{\textbf{FIRE-2 ($z=1$)}} 
        & \centering Young (face-on)          & 56.0 & 266 & 795  & 3420 \\
        & \centering Old (face-on)            & 23.4  & 79.7 & 107 & 240 \\
\enddata
\tablecomments{DMs are in units of pc cm$^{-3}$ and are corrected for redshift suppression. Young populations are $<10\,\mathrm{Myr}$, Old populations are $>10\,\mathrm{Myr}$. FIRE-2 percentiles are reported for simulated MW-mass galaxies.}
\end{deluxetable*}

Notably, this simulation primarily considers face-on galaxies (inclination = 0), with most sightlines beginning from or intersecting a single H\,\textsc{ii} region near the source. Therefore, the elevated DM$_{\text{host}}$ values of the DSA-110 and literature sample are most consistent with a scenario where a single overdense ISM region in the FRB local environment is the dominant source of dispersion from the host galaxy. These local contributions would mask the expected DM-z correlation in Figure~\ref{fig:redshift_trends} associated with the relatively modest (factor of $\sim 5$ at $z = 1.5$) increase in the bulk ISM density with redshift \citep{kaasinen2017,kovacs2024}. This link between FRBs and star-forming regions is consistent with the findings of \citet{sharma2024}, which showed that FRBs preferentially occur in star-forming galaxies, using the same DSA-110 sample. 

This evidence favors a young progenitor model for FRBs, with emission occurring less than 10 Myr after star formation. We also consider whether these large estimates of DM$_{\text{host}}$ might be dominated by even more concentrated environments, such as supernova remnants and other compact nebulae, rather than more extended H\,\textsc{ii} regions. Requiring the free-free opacity of the medium to remain below one at GHz frequencies places a lower limit on the path length through the dense ionized gas (Law et al., in prep.). Under permissive assumptions for the gas temperature and clumping, we find that path lengths as short as $R_\mathrm{pc} \gtrsim 0.1\ \mathrm{pc}$ can reproduce the observed DMs while remaining transparent to FRB emission. Therefore, this constraint does not exclude environments such as young supernova remnants or magnetar wind nebulae as viable sources of the excess DM. However, such dynamic local environments are generally disfavored for the FRB population as a whole by their modest RMs \citep{sherman2024}, the lack of significant DM or scattering variability in most repeaters \citep{curtin2025}, and the absence of extreme pulse broadening \citep{ravi2019}.

\subsection{The $\tau$--DM$_{\text{host}}$ Correlation for FRBs}
\label{sec:tau_dm}

Using measurements of pulse broadening, we investigate the dominant sources of extragalactic scattering to further constrain the local environment of FRBs. We combine the DSA-110 sample with an additional set of FRBs from the literature with measured redshifts and scattering times, drawn from \citet{cordes2022} and references therein. Scattering timescales from the literature sample were primarily obtained using direct measurements of pulse broadening, with the exception of FRB 20200120E, for which a nanosecond-scale timescale was inferred from a measured scintillation bandwidth using the relation $\Delta\nu_{\mathrm{d}} \simeq (2\pi \tau)^{-1}$ \citep{nimmo2022}. 

\begin{figure}[t!]
    \centering
    \includegraphics[width=0.475\textwidth]{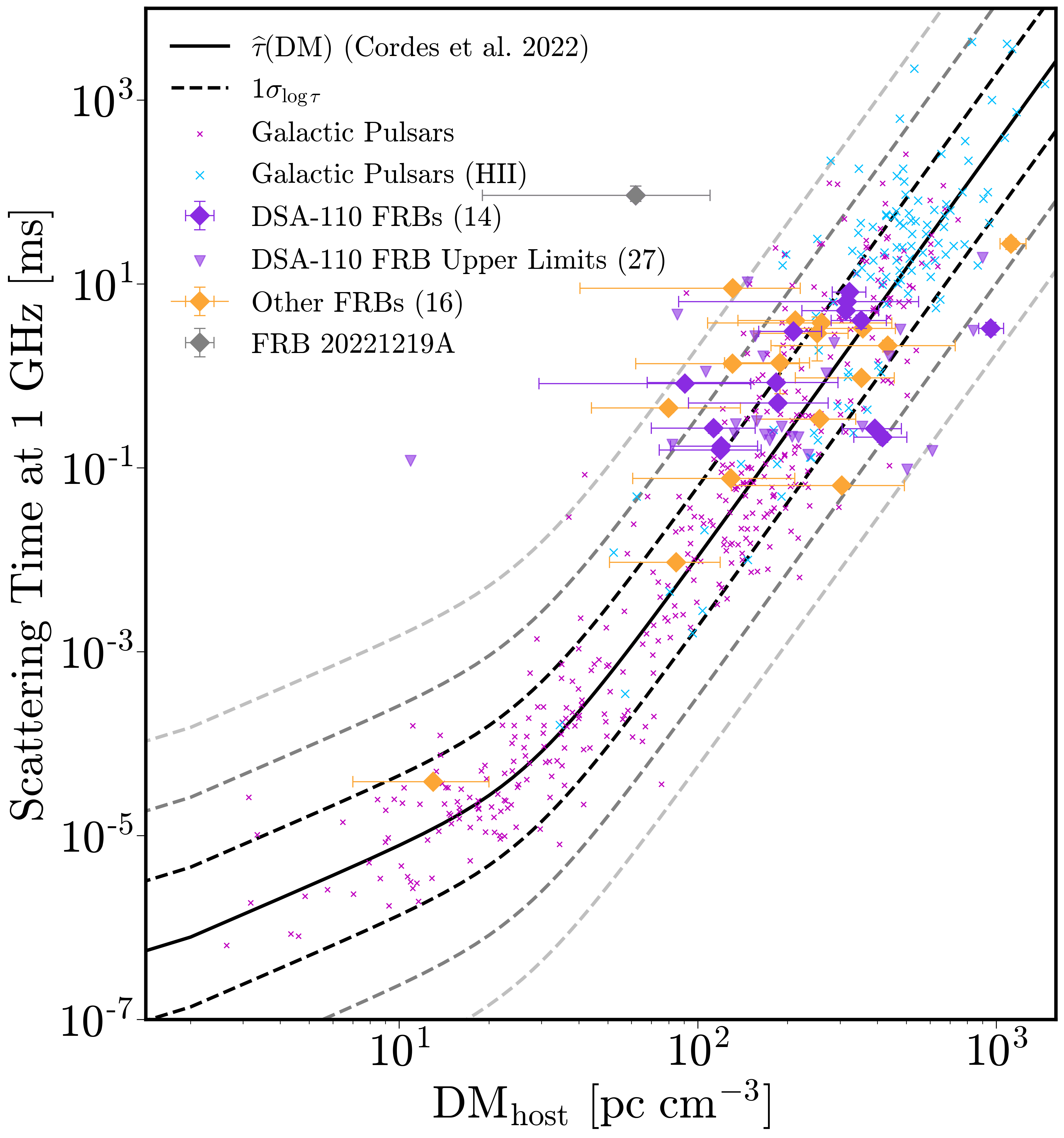}
    \caption{\(\tau\)--DM relation for Galactic pulsars and FRBs. The fitted line (solid black) and \(\pm1\sigma\) variations (dashed black) are based on measurements and upper limits on \(\tau\) for Galactic pulsars (568 in total; data provided courtesy of J. Cordes and published in \citep{cordes2016}. The subset of Galactic pulsars with sightlines intersecting known H\,\textsc{ii} regions are shown in blue \citep{ocker2024}. The scattering timescales and \(\text{DM}_{\text{host}}\) estimates for the subset of well-localized and scattered DSA-110 FRBs are over-plotted in purple. A heavily scattered DSA-110 FRB is shown in gray, with scattering attributed to the CGM of an intervening galaxy \citep{faber2024}. An additional sample of FRBs are shown in orange, drawn from \citet{cordes2022} and references therein. All FRB scattering times have been scaled to their nominal values at 1 GHz in the host-galaxy rest frame, and by a factor of 1/3 to compare with pulsar scattering.}
    \label{fig:hockey_stick}
\end{figure}

We compare this combined FRB sample to an established empirical $\tau$--DM relation for Galactic pulsars. Most recently, this relation was derived by fitting scattering timescales and dispersion measures (DMs) for 568 Galactic pulsars using the canonical fitting function
\[
\tau_{\mathrm{b}}(\mathrm{DM}) = A \times \mathrm{DM}^{a} \left(1 + B \times \mathrm{DM}^{b} \right),
\]
\citep{ramachandran1997,cordes2022}. This yields the empirical form

\begin{equation}
    \begin{aligned}
    \left[\widehat{\tau}(\mathrm{DM}, \nu)\right]_{\mathrm{mw,psr}} = 1.90 \times 10^{-7}~\mathrm{ms} \times \nu^{-\alpha} \mathrm{DM}^{1.5}
    \\
    \times \left(1 + 3.55 \times 10^{-5} \mathrm{DM}^{3.0} \right)
    \end{aligned}
    \label{fig:pulsar_model}
\end{equation}
where $\nu$ is the observing frequency in GHz. The relation exhibits a scatter of $\sigma_{\log \tau} = 0.76$ dex \citep{cordes2022}. Figure~\ref{fig:hockey_stick} shows this $\tau$--DM relation for Galactic pulsars and the distribution of measured pulsar $\tau$ and DM values. The relation steepens significantly at large DMs due to the larger density fluctuations of the inner Galaxy where high-DM pulsars are found \citep{ocker2021,cordes2022}. The difference between spherical wavefronts from Galactic pulsars and plane waves from extragalactic FRBs amounts to an increase by a factor of 3 in the scattering time \citep{cordes2016}. To enable a one-to-one comparison between pulsars and FRBs, we apply this geometric correction to FRB scattering times. We neglect any contribution to $\tau$ from the Milky Way ISM or halo, given that such an interaction is highly suppressed by the scattering geometry and negligible above 1 GHz \citep{ocker2021,ocker2025}. Assuming that pulse broadening is dominated by a scattering layer within or near the host-galaxy, we scale $\tau$ to the rest frame using the measured host-galaxy redshift by applying a $(1+z)^{3}$ correction factor. This assumption is supported by the two-screen scattering models of the FRBs for which the model could be applied, all of which place the dominant scattering material on galactic scales near each source. Lastly, one FRB with extreme scattering from multiple intervening galaxies (FRB 20221219A) is excluded from the $\tau$–DM$_{\text{host}}$ correlation analysis \citep{faber2024}.

We find a significant $\tau$--DM$_{\text{host}}$ correlation for FRBs by performing a Spearman Rho test ($\rho = 0.52$, $p = 0.003$ for the combined sample of FRBs, $\rho = 0.35$, $p = 0.2$ for DSA-110 FRBs, and $\rho = 0.52$, $p = 0.04$ for the literature sample). Excluding FRB 20200120E from this analysis, an outlier with minimal scattering embedded in a globular cluster, we still recover a correlation ($\rho = 0.47$, $p=0.01$) for the combined FRB sample. Furthermore, we find that the $\tau$--DM$_{\text{host}}$ distribution for FRBs is consistent with the relation established for Galactic pulsar scattering. We test for consistency by performing a $\chi^2$ test in which the uncertainties in DM$_{\text{host}}$ and $\tau$ are marginalized alongside the intrinsic dispersion of the pulsar distribution, taking the null hypothesis that FRBs follow this empirical model. We obtain $\chi_\nu^2 = 1.11$, $p = 0.31$ for the combined sample, $\chi_\nu^2 = 1.37$, $p = 0.16$ for DSA-110 FRBs, and $\chi_\nu^2 = 0.88$, $p = 0.59$ for the literature sample, with no evidence for a deviation from the pulsar relation. Additionally, the DSA-110 sample exhibits a dispersion of $\sigma_{\log \tau} = 1.07$ dex about the pulsar $\tau$--DM$_{\text{host}}$ relation, comparable to the $\sigma_{\log \tau} = 0.76$ dex of the pulsar relation itself. Despite this general agreement, we note that DSA-110 FRBs with robust $\tau$ measurements appear preferentially offset toward slightly longer scattering times, with 11 of 14 bursts lying above the predicted relation (median residual $+0.57$ dex). The sample also exhibits a shallower $\tau$--DM$_{\text{host}}$ correlation than predicted by the pulsar model over the corresponding DM range. Both discrepancies may result from a selection bias introduced by the time finite resolution of the data, which makes narrower FRBs more difficult to detect and their pulse broadening more difficult to measure. This is likely to bias the sample of $\tau$ measurements in favor of FRBs that occupy a more highly scattered regime, independent of DM$_{\text{host}}$, producing the observed trend. This is supported by the similar offset towards higher $\tau$ seen in the DSA-110 upper limits, as well as 14/27 of the upper limits being concentrated around $\tau = 10^{-1}\ \mathrm{ms}$, suggesting that many of these FRBs may have scattering times shorter than can be easily resolved in DSA-110 data, but which may still follow the predicted relation. As exemplified by one DSA-110 FRB (FRB 20221219A, shown in gray) scattered by intervening halos \citep{faber2024}, a scattering medium at a cosmological distance from both the source and the observer greatly enhances $\tau$ and may contribute little DM \citep{cordes2022}. This degree of overscattering is not observed among any other FRBs in the sample, and does not mask the $\tau$--DM$_{\text{host}}$ correlation expected from host-galaxy scattering. Therefore, FRB scattering appears broadly consistent with the pulsar scattering relation, suggesting that $\tau$ is dominated by scattering from density fluctuations in the host-galaxy ISM.

In particular, FRB scattering appears most consistent with the steep portion of the pulsar $\tau$--DM relation that is dominated by the enhanced scattering along sightlines passing through the inner Galaxy and the plane of the Galactic disk. To investigate a potential association between FRBs and H\,\textsc{ii} regions, we compare the sample to the subset of Galactic pulsars with sightlines intersecting foreground H\,\textsc{ii} regions \citep{ocker2024}. Despite the sample likely being biased toward more heavily scattered bursts, we find that FRBs exhibit lower $\tau$ and DM$_{\text{host}}$ values than most pulsars with foreground H\,\textsc{ii} regions. However, Galactic sources with foreground H\,\textsc{ii} regions are not directly analogous to extragalactic sources embedded within H\,\textsc{ii} regions. First, scattering from an intervening screen located near the source is enhanced with respect to scattering from a screen containing the source \citep{cordes2022}. A source embedded within an H\,\textsc{ii} region would therefore exhibit less scattering and less dispersion due to the shorter path length through the overdense region. Moreover, because most H\,\textsc{ii} regions are near the inner Galaxy, and most pulsars in this sample are located behind them in the Galactic disk, we are biased toward long sightlines that traverse both the inner Galaxy and the plane of the disk \citep{ocker2024}. These geometries are not representative of typical FRB sightlines through their host galaxies and may lead to systematically higher DM and $\tau$ values than would be observed for a similar population viewed by an extragalactic observer. Therefore, increased scattering of H\,\textsc{ii} region associated pulsars relative to FRBs is not unexpected, and remains consistent with an FRB population preferentially embedded within regions of high star formation.

\section{Discussion: Implications for FRB Progenitors}
\label{sec:discussion}

The growing number of FRBs localized to host galaxies has opened new avenues for studying their local environments as a population. By constraining how host galaxies contribute to FRB observables, we are now able to perform comparisons with Galactic pulsars and simulations of potential progenitor populations. We interpret these combined observables to assess whether FRBs preferentially occur in star-forming regions, and indeed in H\,\textsc{ii} regions, and discuss implications of these results for both progenitor models and cosmological applications of FRBs. 

An investigation of the DSA-110 sample suggests that FRBs receive large DM contributions from within their host galaxies. The observed DM$_{\text{host}}$ distribution and its redshift dependence are most consistent with simulations of young stellar progenitors ($<10\,\mathrm{Myr}$). These populations of young progenitors are associated with regions of increased star-formation, such as H\,\textsc{ii} regions, with ISM overdensities that are able to contribute the observed DM excess. In contrast, simulations that assume older and more widely distributed progenitors throughout the host galaxy systematically underestimate the observed DM$_{\text{host}}$ values. We also find a significant correlation between RM$_{\text{host}}$ and DM$_{\text{host}}$, suggesting that FRB RMs are typically dominated by contributions from the host-galaxy ISM (including, e.g., H\,\textsc{ii} regions), rather than dynamic local environments. This correlation supports the use of RM$_{\text{host}}$ as a probe of large-scale magnetic fields in most FRB environments. Additionally, the inferred line-of-sight magnetic field strengths for FRBs in the DSA-110 sample are characteristically larger than those of the Milky Way ISM, but generally much weaker than the extreme fields observed in the dynamic local environments of some repeaters. While direct measurements of field strengths in H\,\textsc{ii} regions are difficult, the elevated values of some DSA-110 FRBs appear broadly consistent with estimates of typical magnetism in Galactic H\,\textsc{ii} regions. 

The extragalactic scattering of FRBs is also consistent with the host-galaxy ISM being the dominant scattering medium. FRBs in a combined DSA-110 and literature sample follow a $\tau$--DM$_{\text{host}}$ relation similar to that observed for Galactic pulsars, suggesting that they typically occupy a scattering regime primarily shaped by diffuse interstellar turbulence rather than extreme local environments. Specifically, FRBs occupy a scattering regime most comparable to pulsars whose sightlines intersect the more turbulent plasma near the inner Galaxy. This similarity may further support a scenario in which the FRB population is predominantly embedded in active star-forming environments with large ISM density fluctuations. 

Taken together, the inferred gas densities, magnetic fields, and density fluctuations within FRB local environments appear to favor a scenario in which FRB progenitors preferentially form within regions of elevated star formation and remain embedded in these environments during their emission lifetime. This supports source models with short formation-delay times post star formation, such as magnetars formed via core-collapse supernova (CCSNe), as the dominant pathway for FRB production. Assuming that FRBs are young remnants of CCSNe and remain embedded within or near H\,\textsc{ii} regions during their emission lifetime, we can place rough constrains on their typical kick velocities. For a typical H\,\textsc{ii} region of scale height $\sim 50\ \mathrm{pc}$ and a characteristic FRB lifetime of $\sim 5\ \mathrm{Myr}$ (consistent with magnetar ages) FRB progenitors must have typical 1D rms kick velocities $\sigma <1000\,\mathrm{km\,s^{-1}}$. This is comparable to the observed distribution of pulsar kick velocities, which have a mean $\sigma = 265\,\mathrm{km\,s^{-1}}$, and rarely exceed $1000\,\mathrm{km\,s^{-1}}$ \citep{hobbs2005}. Recent results suggest that the characteristic kick velocities of pulsars may be lower by as much as $\sim 50\%$, but this rough upper limit for the FRB population remains consistent with this updated distribution \citep{disberg2025}. 

Direct localization of FRBs to specific galactic structures, such as H\,\textsc{ii} regions or spiral arms, is only feasible for nearby events ($z\lesssim0.1$) with sub-arcsecond precision. To date, only a small number of bursts have been localized with sufficient precision to identify these potential associations \citep{mannings2021,woodland2024,gordon2025,abbott2025,blanchard2025}. The population-level analysis of burst properties presented in this paper offers a complementary approach to constraining FRB local environments. As the number of well-localized FRBs detected by DSA-110 and other interferometric surveys continues to grow, the distribution of observed burst properties will serve as a powerful statistical probe of FRB formation sites. Despite growing evidence linking FRBs to galaxies and regions with high star formation \citep{gordon2023,sharma2024}, targeted searches of nearby star forming galaxies have thus far not yielded any detections \citep{paine2024,wharton2025}. Furthermore, the localization of some FRBs to less-active environments and galaxies with low star formation rates (e.g., FRB 20210117A \citealt{bhandari2023}, FRB 20200120E \citealt{kirsten2022}, FRB 20240214A \citealt{shah2025}, and several sources in \citealt{gordon2023}) may indicate that a subset of the population forms through more delayed channels. Continued efforts to localize FRBs will be essential to determining whether most progenitors originate in young stellar environments, and if they share a dominant formation pathway with pulsars and magnetars. In particular, expanding the sample of well-localized high-redshift FRBs will be crucial for tracing redshift dependent effects, such as the predicted changes in the distribution of DM$_{\text{host}}$ as the galaxy population evolves \citep{orr2024}. 

Improved characterization of burst properties will also enhance their utility as cosmological probes. An ongoing uncertainty in the application of FRBs to studying the IGM is the poorly constrained contribution to DM from the host galaxy. While cosmological studies often adopt a fiducial value of DM$_{\mathrm{host}} \sim 50-100\ \mathrm{pc \, cm^{-3}}$  \citep{arcus2021}, our results suggest that this contribution may be significantly underestimated, with typical values several times larger than commonly assumed. By attributing a larger fraction of the observed DM in unlocalized sources to DM$_{\mathrm{host}}$, rather than DM$_{\mathrm{IGM}}$, we favor the conclusion of \citet{orr2024} that the population of unlocalized FRBs may be at lower redshifts than previously inferred. We define the extragalactic DM contribution as
 \begin{equation}
    \text{DM}_{\text{exgal}}(z) = \text{DM}_{\text{IGM}}(z) + \frac{\text{DM}_{\text{host}}}{(1+z)}
    \label{eq:dm_budget}
\end{equation}
where $\text{DM}_{\text{IGM}}(z)$ takes the form described by \citet{connor2025}. Using this, we derive a rough estimate of $\text{DM}_{\text{exgal}}(z)$ for our inferred median rest-frame DM$_{\mathrm{host}} = 242.2\ \mathrm{pc \, cm^{-3}}$, and compare this to the result obtained under a more traditional assumptions of DM$_{\mathrm{host}} \approx 50-100\ \mathrm{pc \, cm^{-3}}$. From this comparison, we find that the redshift estimates of unlocalized FRBs with $\mathrm{DM}_{\mathrm{exgal}}<500\ \mathrm{pc \, cm^{-3}}$ could be reduced by $\Delta z \sim -0.1$. Therefore, continuing to robustly characterize the distribution of DM$_{\mathrm{host}}$ across the FRB population will be necessary to use the large sample of unlocalized FRBs for precise cosmological analyses and for refining estimates of the cosmic baryon distribution.

\vspace{15mm}

\section{Conclusion}
\label{sec:conclusion}

For the sample of 43 FRBs detected with the DSA-110 and precisely localized to their host galaxies, we have isolated the host-galaxy contributions to DMs, RMs, and scattering timescales. Motivated by what these propagation effects reveal about the local environments of FRBs, we have analyzed the DSA-110 sample alongside other FRBs drawn from the literature to characterize these effects across the population. From this analysis, we conclude the following regarding the source environments and potential progenitors of FRBs: 

\begin{itemize}
    \renewcommand\labelitemi{---}
    \item  FRB DMs contain significantly larger host-galaxy contributions than previously assumed, with a median inferred rest-frame DM$_{\mathrm{host}}$ of 241 $\mathrm{pc \, cm^{-3}}$ for the DSA-110 sample. This excess DM indicates that FRBs tend to trace overdense regions of ionized gas within the host-galaxy ISM, such as H\,\textsc{ii} regions. 

    \item DM$_{\mathrm{host}}$ and RM$_{\mathrm{host}}$ are significantly correlated across our combined DSA-110 and literature sample of FRBs, implying that the host-galaxy ISM is also the dominant source of RMs in most FRBs. 

    \item The line-of-sight magnetic field strengths within FRB host galaxies are characteristically $1–2 \mu\text{G}$ stronger than those probed by Galactic pulsars. Comparison of FRBs with Galactic pulsars viewed through foreground H\,\textsc{ii} regions reveals that FRBs may trace environments with similarly enhanced magnetism, consistent with an association with star-forming regions in their host galaxies. 

    \item Across the redshift range $0 < z < 1.33$, we find no significant evolution of DM$_{\mathrm{host}}$, RM$_{\mathrm{host}}$, or ${B}_{\parallel, \text{host}}$. We suggest that the potential association of FRBs with more active star-forming environments could limit the extent to which the population traces overall changes in the density or magnetism of the broader host-galaxy ISM.
    
    \item FRB scattering timescales and DM$_{\mathrm{host}}$ are strongly correlated and follow a relationship consistent with that observed for Galactic pulsars, indicating that diffuse turbulence in the host-galaxy ISM is the primary source of extragalactic scattering. Using a two-screen scattering model, combined measurements of pulse broadening and scintillation constrain the dominant extragalactic scattering medium to lie near or within the host galaxy of most FRB sources. 
    
    \item The inferred distribution of DM$_{\mathrm{host}}$ for FRBs and its lack of direct redshift dependence are most consistent with predictions from cosmological simulations of young ($<10\,\mathrm{Myr}$) progenitor populations. These findings favor magnetars formed in core-collapse supernovae as the dominant source of FRBs. 
\end{itemize}

\vspace{5mm}

\section{Acknowledgments}

The authors thank staff members of the Owens Valley Radio Observatory and the Caltech radio group, whose efforts were instrumental to the success of the DSA-110. The DSA-110 is supported by the National Science Foundation Mid-Scale Innovations Program in Astronomical Sciences (MSIP) under grant AST-1836018. This material is based upon work supported in part by the National Science Foundation Graduate Research Fellowship under Grant No. 2139433.

\clearpage

\bibliographystyle{aasjournal}
\bibliography{verdi2025}

\appendix
\twocolumngrid
\renewcommand{\thetable}{A\arabic{table}}
\setcounter{table}{0}

\renewcommand{\thefigure}{A\arabic{figure}}
\setcounter{figure}{0}

\begin{deluxetable}{ccccc}[t!]
\tablecaption{DSA-110 FRB Localizations.\label{tab:frbs}}
\tablewidth{0pt}
\tablehead{
\colhead{FRB Name} & \colhead{R.A.} & \colhead{Decl.} & \colhead{R.A. Error} & \colhead{Decl. Error} \\
\colhead{} & \colhead{(J2000)} & \colhead{(J2000)} & \colhead{(arcsec.)} & \colhead{(arcsec.)} 
}
\startdata
20230913G & 20$^{\rm h}$20$^{\rm m}$08$^{\rm s}$.92 & $+$70$^{\circ}$47\arcmin33\arcsec.96 & 0.8 & 0.5 \\
20240104A & 23$^{\rm h}$15$^{\rm m}$29$^{\rm s}$.76 & $+$72$^{\circ}$49\arcmin14\arcsec.1 & 1.3 & 0.9 \\
20240203D & 20$^{\rm h}$50$^{\rm m}$28$^{\rm s}$.59 & $+$73$^{\circ}$54\arcmin00\arcsec.0 & 0.8 & 0.6 \\
20240224A & 04$^{\rm h}$40$^{\rm m}$27$^{\rm s}$.57 & $+$73$^{\circ}$30\arcmin46\arcsec.0 & 0.6 & 0.5 \\
\enddata
\end{deluxetable}

In this work, we present the DSA-110 detections and interferometric localizations of four FRBs for which new redshift measurements of the host galaxies were obtained. Localizations are given in Table~\ref{tab:frbs}, with further details reported to the Transient Name Server. 

\begin{figure}[t!]
\centering 
    \includegraphics[width=0.23\textwidth]{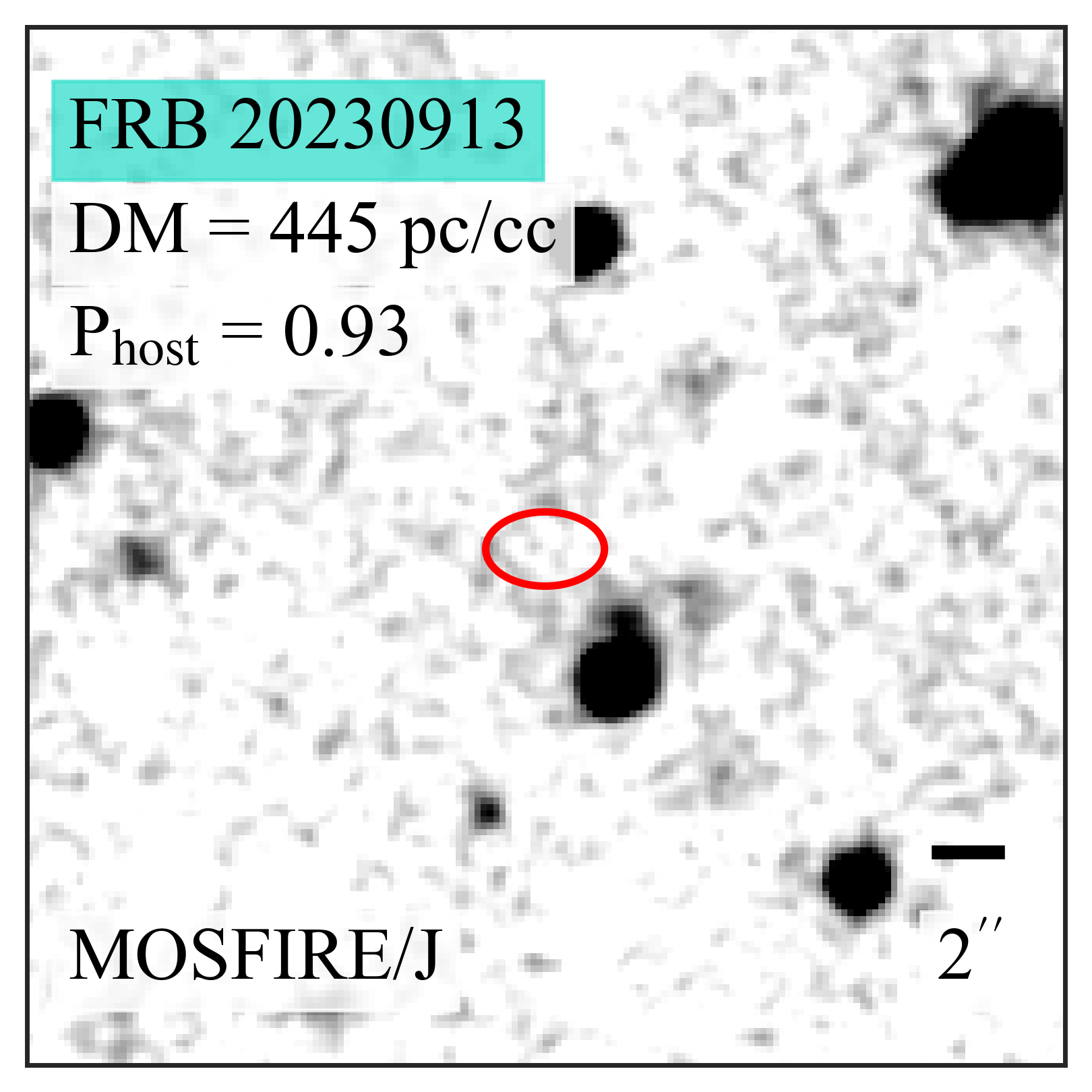}
    \includegraphics[width=0.23\textwidth]{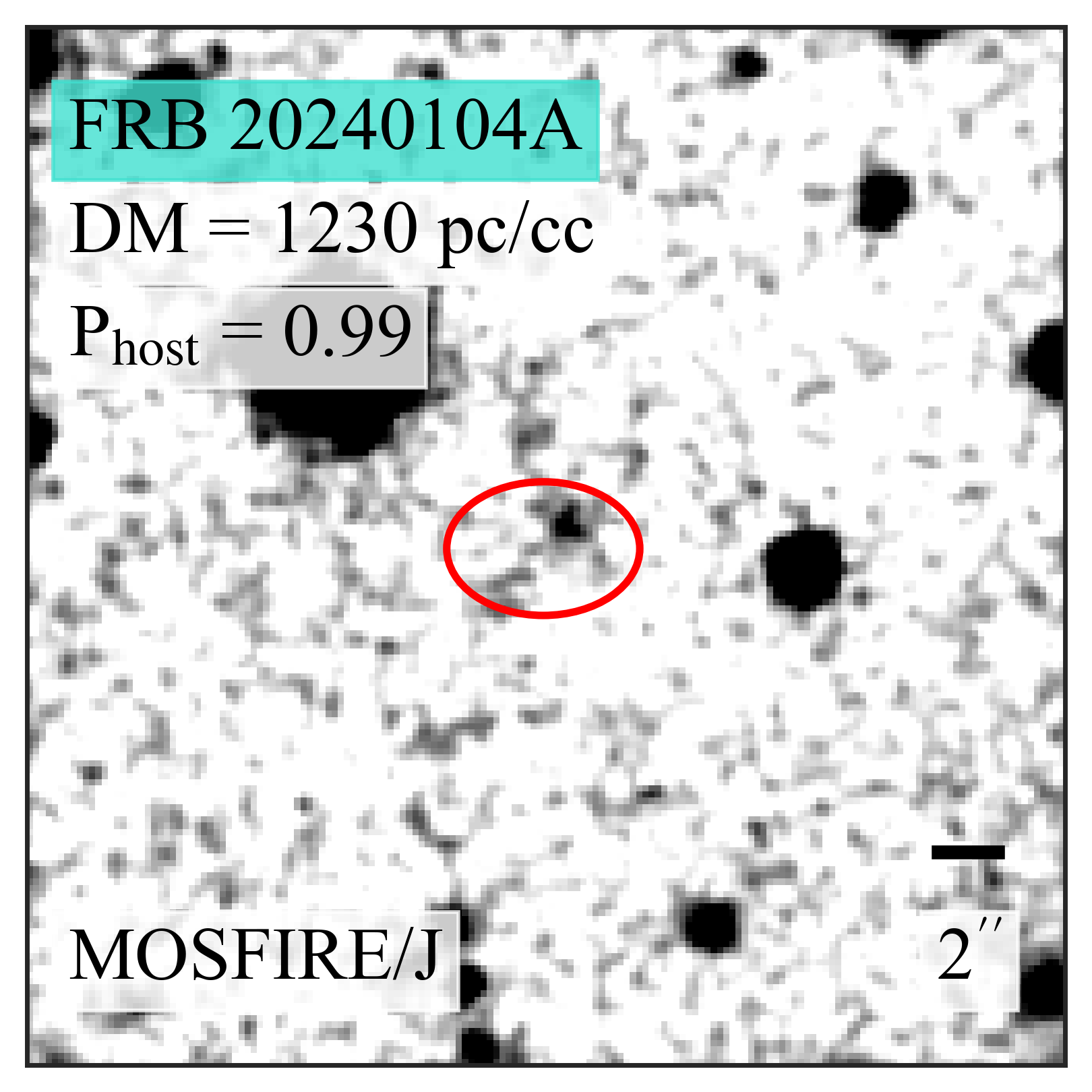}
    \includegraphics[width=0.23\textwidth]{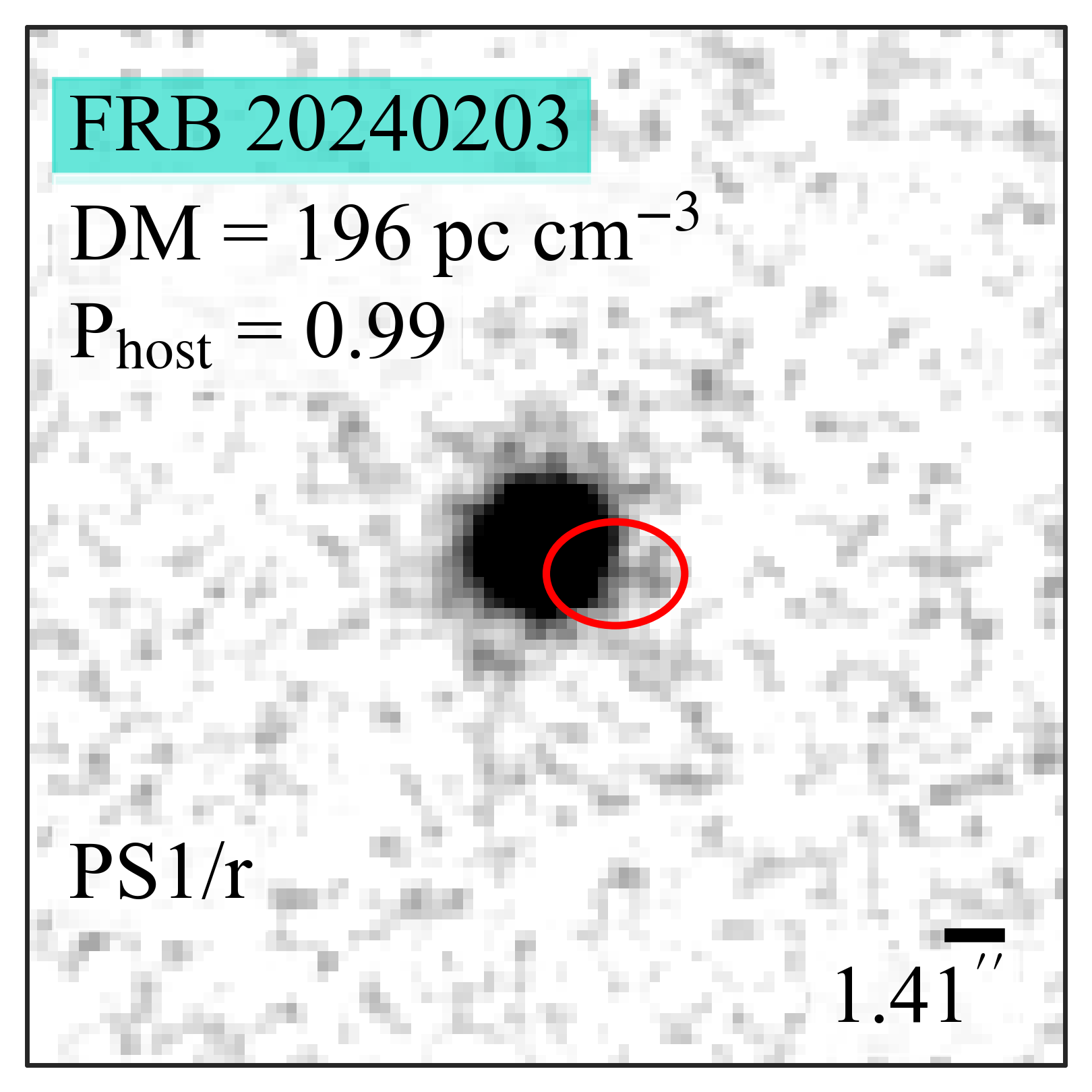}
    \includegraphics[width=0.23\textwidth]{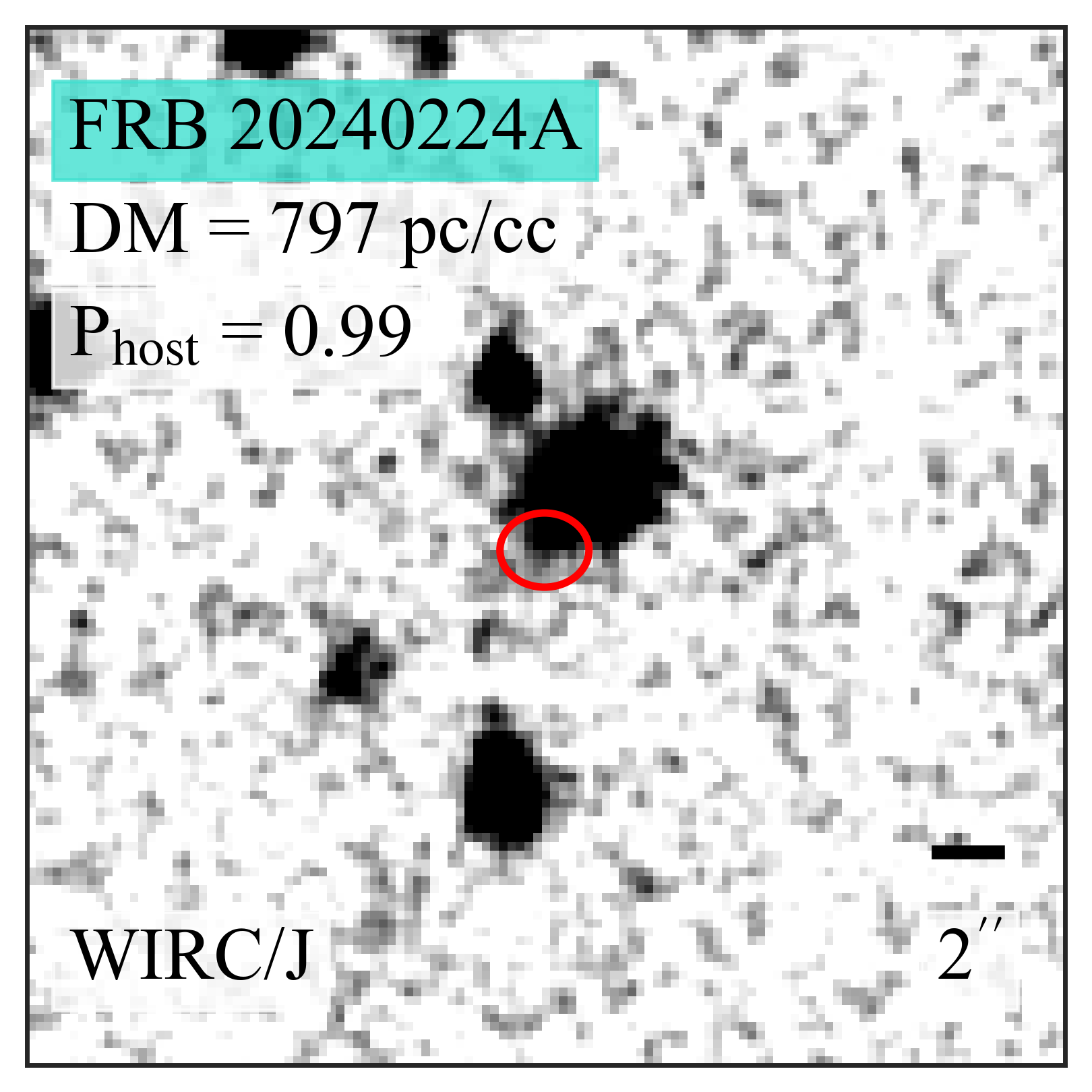}
    \caption{Images of the likely host galaxies of four DSA-110 FRBs for which spectroscopic redshifts were obtained, together with the 90\% confidence localization ellipses. Host-association probabilities from the \texttt{astropath} software software are indicated, together with the source of the imaging data.}
    \label{fig:frbhosts}
\end{figure}

These four FRBs had host galaxies identified via various archival and new optical/IR imaging programs, with associations established with the \texttt{astropath} software \citep{astropath} with probabilities of $>90\%$. Images of the hosts and their localizations are shown in Figure~\ref{fig:frbhosts}. 

\begin{figure*}
\centering 
    \includegraphics[width=0.54\textwidth]{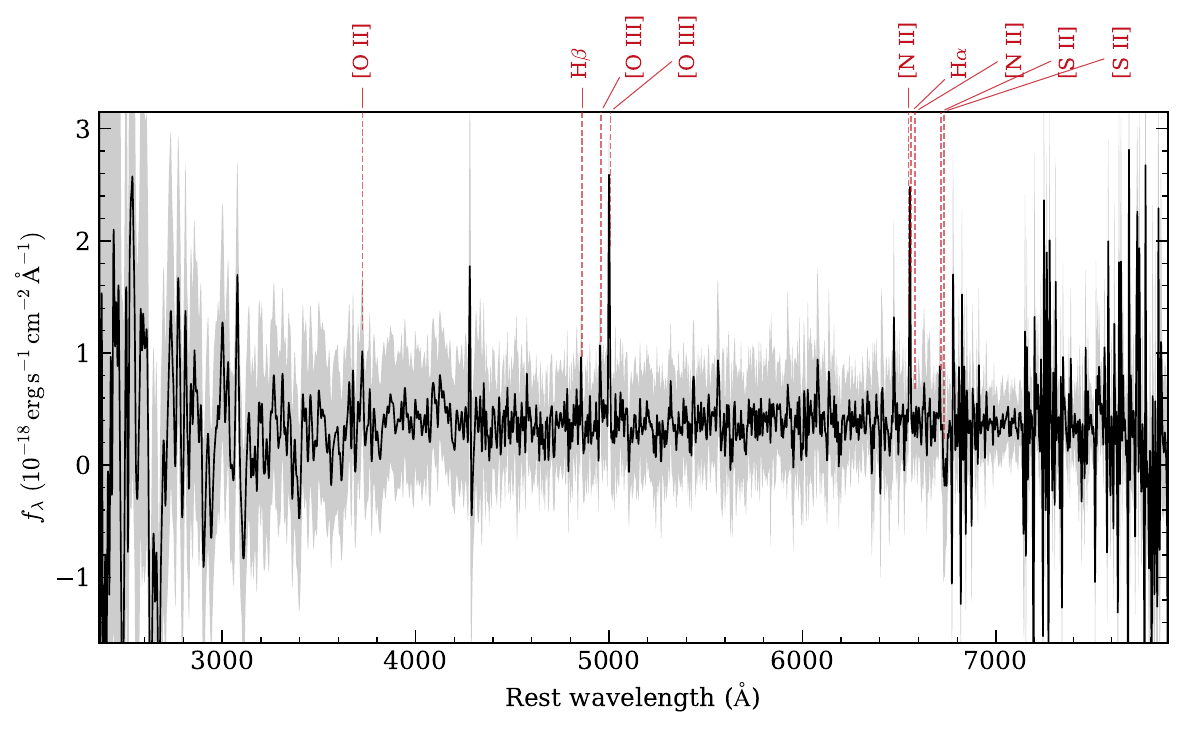} \\
    \includegraphics[width=0.54\textwidth]{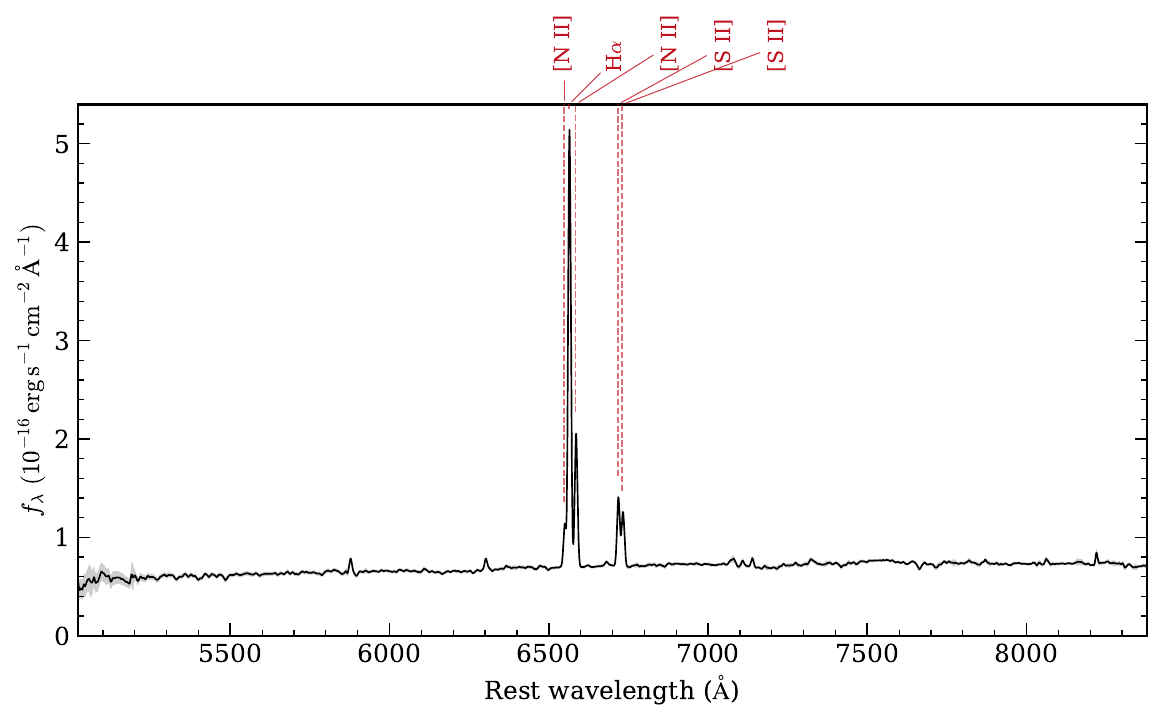}  \\
    \includegraphics[width=0.54\textwidth]{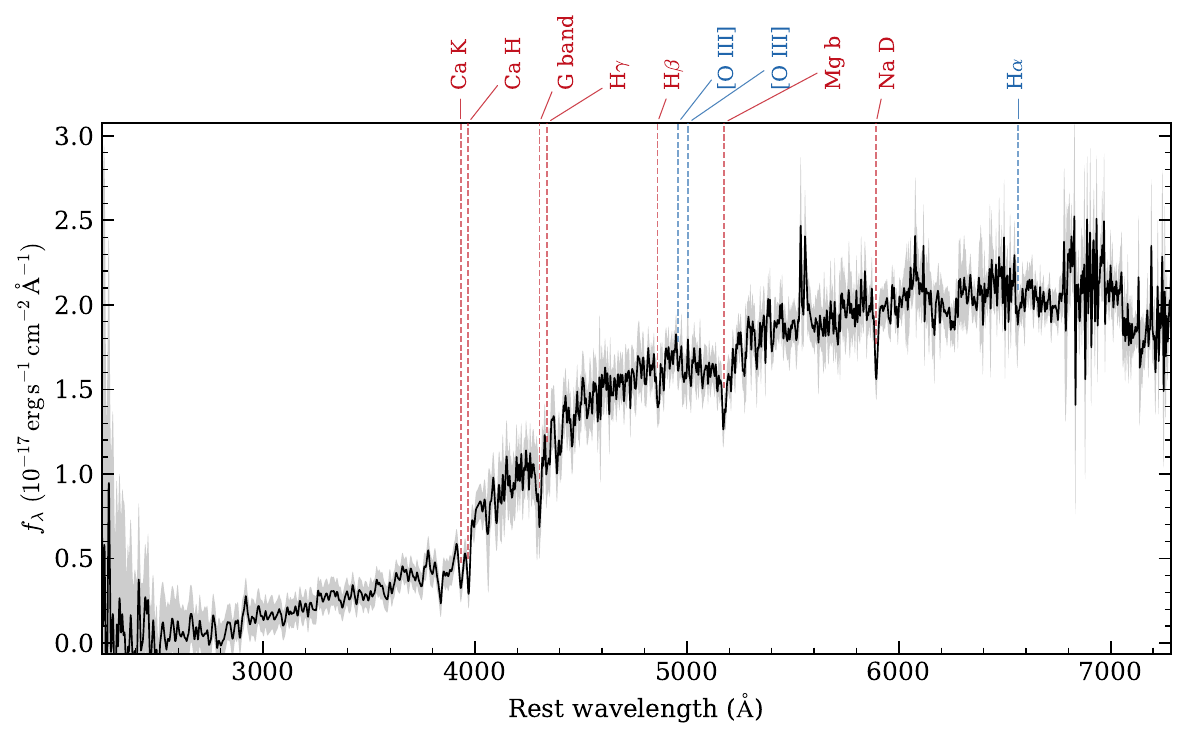}  
    \caption{Optical spectroscopy of the host galaxies of FRB\,20230913G (top, $z=0.3024$),  FRB\,20240203D (middle, $z=0.074$), and FRB\,20240224A (bottom, $z=0.37$). Spectra were obtained with Keck-I/LRIS. The redshifts were measured using the marked emission lines. These spectra are not corrected for galactic dust-extinction and slit losses. }
    \label{fig:otherfrbspec}
\end{figure*}

\textbf{FRB\,20230913G:} We observed the host galaxy on 2025 May 26 with Keck-I/LRIS, with an exposure time of 5400\,s. The observing setup and reduction were identical to the observations described above. Emission lines were clearly detected (Figure~\ref{fig:otherfrbspec}) at a redshift of $z=0.3024$. 

\textbf{FRB\,20240104A:} We obtained optical spectroscopy of the faint host galaxy with the W.~M.~Keck Observatory Low Resolution Imaging Spectrometer \citep[Keck-I/LRIS;][]{lris} on 2024 October 7. A total of 12600\,s of exposure with a 1\arcsec~slit was recorded using the D560 dichroic, the 400/3400 grism, and the 400/8500 grating at a central wavelength of 7830\,\AA. Data were reduced with \texttt{lpipe} \citep{lpipe} A single emission line was detected (Figure~\ref{fig:mike}), which given the high DM (1351\,pc\,cm$^{-3}$) was tentatively guessed to be the [OII] ($\lambda \lambda $ 3727,3729) doublet at a redshift of $z=1.33$. On 2024 October 20, we then obtained 3600\,s of exposure on the host galaxy with the Multi-Object Spectrometer For Infra-Red Exploration \citep[Keck-I/MOSFIRE;][]{mosfire} using a 0.7\arcsec~long slit in the H band. Data were reduced with \texttt{PypeIt} \citep{pypeit}. Emission lines corresponding to H${\rm \alpha}$ and the [SII] ($\lambda \lambda $ 6716, 6731) doublet were observed, confirming the redshift. 

\textbf{FRB\,20240203D:} We observed the host galaxy on 2024 June 10 with Keck-I/LRIS, with an exposure time of 1800\,s. The observing setup and reduction were identical to the observations described above. Several emission lines were clearly detected (Figure~\ref{fig:otherfrbspec}) at a redshift of $z=0.074$. 

\textbf{FRB\,20240224A:} We observed the host galaxy on 2025 December 17 with Keck-I/LRIS, with an exposure time of 1800\,s. The observing setup and reduction were identical to the observations described above. Several absorption lines were clearly detected (Figure~\ref{fig:otherfrbspec}) at a redshift of $z=0.3726$. Notably, no emission lines were detected.

\begin{deluxetable*}{ccccccccccc}[t]
\label{tab:burst_props}
\tabletypesize{\scriptsize}
\renewcommand{\arraystretch}{1.24}
\tablewidth{0pt}
\tablecaption{DSA-110 FRB Host-Galaxy Redshifts, Burst Properties, and Scattering\label{table:burst_props}}
\tablehead{FRB & Redshift & DM & DM$_{\mathrm{host}}$ & RM & RM$_{\mathrm{host}}$ & $\hat{B}_{\parallel,\mathrm{host}}$ ($\mu$G) & $\tau_{\mathrm{1\,GHz}}$ (ms) & $\Delta\nu$ (kHz) & $L_x L_g$ (kpc$^2$) & $b_{\mathrm{gal}}\ (^{\circ})$}
\startdata
20220204A & 0.4000 & $612.20$ & $269.75^{+109.15}_{-113.95}$ & $10.95 \pm 11.71$ & $-4.54 \pm 30.34$ & $-0.03^{+0.11}_{-0.29}$ & $<1.167$ & -- & -- & 28.3 \\
\hline
20220207C & 0.0430 & $262.30$ & $119.98^{+38.98}_{-38.96}$ & $-162.48 \pm 0.04$ & $-174.24 \pm 18.42$ & $-2.16^{+0.67}_{-0.53}$ & $0.454^{+0.014}_{-0.013}$ & $1029 \pm 65$ & -$\lesssim 22.5$ & 18.4 \\
\hline
20220208A & 0.3510 & $437.00$ & $85.52^{+60.29}_{-60.33}$ & $23.27 \pm 9.70$ & $44.18 \pm 33.29$ & $1.58^{+0.56}_{-1.45}$ & $<5.663$ & -- & -- & 13.6 \\
\hline
20220307B & 0.2507 & $499.15$ & $174.34^{+75.51}_{-78.78}$ & $947.23 \pm 12.27$ & $1486.00 \pm 48.74$ & $12.79^{+4.65}_{-5.57}$ & $<0.307$ & -- & -- & 10.5 \\
\hline
20220310F & 0.4790 & $462.15$ & $82.32^{+62.15}_{-60.91}$ & $-11.39 \pm 0.19$ & $9.00 \pm 16.26$ & $0.44^{+0.19}_{-0.61}$ & $<0.167$ & -- & -- & 34.8 \\
\hline
20220319D & 0.0111 & $110.95$ & $10.91^{+9.07}_{-8.89}$ & $-59.94 \pm 14.33$ & $-47.30 \pm 23.26$ & $-9.84^{+7.80}_{-8.44}$ & $<0.348$ & -- & -- & 9.1 \\
\hline
20220330D & 0.3714 & $468.10$ & $146.90^{+83.17}_{-86.45}$ & $122.06 \pm 4.54$ & $270.16 \pm 17.79$ & $3.87^{+1.11}_{-2.50}$ & $<12.130$ & -- & -- & 43.7 \\
\hline
20220418A & 0.6220 & $623.45$ & $132.68^{+94.80}_{-94.40}$ & $-6.13 \pm 7.48$ & $-37.73 \pm 41.67$ & $-0.95^{+0.90}_{-0.51}$ & $<0.167$ & -- & -- & 44.5 \\
\hline
20220506D & 0.3005 & $396.93$ & $90.64^{+60.12}_{-61.19}$ & $32.38 \pm 3.60$ & $63.15 \pm 25.62$ & $1.97^{+0.39}_{-1.61}$ & $1.125^{+0.128}_{-0.120}$ & -- & --& 16.5 \\
\hline
20220509G & 0.0894 & $269.50$ & $119.27^{+44.14}_{-44.93}$ & $109.00 \pm 1.17$ & $109.29 \pm 18.00$ & $1.51^{+0.32}_{-0.88}$ & $0.362^{+0.025}_{-0.033}$ & $3003 \pm 244$ & $\lesssim 363.0$ & 25.5 \\
\hline
20220726A & 0.3610 & $686.55$ & $355.80^{+109.24}_{-110.71}$ & $-499.80 \pm 7.20$ & $-828.56 \pm 37.72$ & $-3.48^{+1.11}_{-0.70}$ & $<0.335$ & -- & -- & 16.3 \\
\hline
20220825A & 0.2414 & $651.20$ & $416.14^{+86.29}_{-82.74}$ & $-750.23 \pm 6.67$ & $-1153.22 \pm 27.32$ & $-3.76^{+0.84}_{-0.56}$ & $0.339^{+0.046}_{-0.042}$ & $8120 \pm 609$ & $\lesssim 5717.0$ & 17.8 \\
\hline
20220831A & 0.2620 & $1146.25$ & $958.35^{+100.64}_{-84.53}$ & $-772.09 \pm 7.19$ & $-1225.45 \pm 33.70$ & $-1.66^{+0.18}_{-0.22}$ & $4.896^{+0.805}_{-0.733}$ & -- & -- & -10.6 \\
\hline
20220914A & 0.1138 & $631.05$ & $503.20^{+54.69}_{-47.78}$ & -- & -- & -- & $<0.209$ & -- & -- & 26.1 \\
\hline
20220920A & 0.1585 & $315.00$ & $134.16^{+54.45}_{-56.40}$ & $830.25 \pm 8.29$ & $1111.31 \pm 20.83$ & $12.23^{+4.11}_{-5.11}$ & $<0.579$ & -- & -- & 38.9 \\
\hline
20221012A & 0.2840 & $442.20$ & $165.67^{+78.09}_{-81.95}$ & $-165.70 \pm 17.66$ & $-280.92 \pm 35.62$ & $-3.19^{+1.80}_{-1.01}$ & $<2.314$ & -- & -- & 26.1 \\
\hline
20221027A & 0.5420 & $452.50$ & $58.64^{+47.33}_{-45.69}$ & -- & -- & -- & -- & -- & -- & 34.2 \\
\hline
20221029A & 0.9750 & $1391.05$ & $837.55^{+298.65}_{-307.13}$ & $155.76 \pm 9.67$ & $674.21 \pm 48.45$ & $1.37^{+0.13}_{-0.68}$ & $<1.211$ & -- & -- & 37.2 \\
\hline
20221101B & 0.2395 & $490.70$ & $167.75^{+73.31}_{-76.62}$ & $32.18 \pm 9.10$ & $69.75 \pm 42.61$ & $0.83^{+0.19}_{-0.82}$ & $<0.365$ & -- & -- & 10.2 \\
\hline
20221113A & 0.2505 & $411.40$ & $122.77^{+66.61}_{-69.65}$ & $-14.35 \pm 10.53$ & $48.22 \pm 32.96$ & $0.95^{+0.23}_{-0.97}$ & $<0.244$ & -- & -- & 15.8 \\
\hline
20221116A & 0.2764 & $640.60$ & $314.07^{+90.41}_{-90.74}$ & $74.04 \pm 11.55$ & $110.49 \pm 32.48$ & $0.56^{-0.02}_{-0.42}$ & $7.404^{+1.202}_{-1.186}$ & -- & -- & 10.0 \\
\hline
20221219A & 0.5540 & $706.70$ & $255.87^{+129.91}_{-136.14}$ & -- & -- & -- & $73.758^{+10.372}_{-10.372}$ & -- & -- & 33.5 \\
\hline
20230124A & 0.0940 & $590.00$ & $477.74^{+48.87}_{-42.85}$ & $1084.47 \pm 12.93$ & $1293.90 \pm 21.03$ & $3.44^{+0.21}_{-0.59}$ & $<7.305$ & -- & -- & 41.0 \\
\hline
20230216A & 0.5310 & $828.00$ & $438.22^{+147.86}_{-151.49}$ & $-339.88 \pm 12.52$ & $-821.83 \pm 33.99$ & $-2.95^{+1.13}_{-0.67}$ & $<1.363$ & -- & -- & 48.0 \\
\hline
20230307A & 0.2710 & $610.15$ & $392.09^{+89.58}_{-87.35}$ & $-476.67 \pm 0.03$ & $-761.65 \pm 12.14$ & $-2.71^{+0.61}_{-0.39}$ & $0.390^{+0.018}_{-0.012}$ & -- & -- & 44.7 \\
\hline
20230501A & 0.3010 & $533.70$ & $177.98^{+83.16}_{-87.32}$ & $-119.70 \pm 1.56$ & $-196.99 \pm 44.04$ & $-2.16^{+1.29}_{-0.51}$ & $<0.313$ & -- & -- & 10.8 \\
\hline
20230521B & 1.3540 & $1345.65$ & $286.60^{+216.79}_{-212.59}$ & $21.00 \pm 13.67$ & $224.50 \pm 191.74$ & $2.36^{+1.12}_{-2.29}$ & $<0.527$ & -- & -- & 9.5 \\
\hline
20230626A & 0.3270 & $451.20$ & $157.87^{+81.10}_{-85.15}$ & $-576.71 \pm 12.42$ & $-1015.47 \pm 31.09$ & $-10.54^{+5.30}_{-4.72}$ & $<0.414$ & -- & -- & 40.0 \\
\hline
20230628A & 0.1265 & $345.15$ & $191.21^{+51.49}_{-51.07}$ & $-17.39 \pm 2.24$ & $-6.18 \pm 9.19$ & $-0.06^{+0.12}_{-0.28}$ & $<0.591$ & -- & -- & 42.7 \\
\hline
20230712A & 0.4525 & $586.96$ & $207.36^{+106.51}_{-111.72}$ & $-77.75 \pm 17.72$ & $-134.76 \pm 40.33$ & $-1.46^{+0.96}_{-0.44}$ & $<0.215$ & -- & -- & 42.5 \\
\hline
20230814B* & 0.5535 & $696.35$ & $183.26^{+112.16}_{-115.52}$ & $18.26 \pm 0.38$ & $20.19 \pm 43.01$ & $0.32^{+0.11}_{-0.69}$ & $0.679^{+0.040}_{-0.041}$ & $582 \pm 72$ & $\lesssim 732.4$ & 13.2 \\
\hline
20230913G & 0.3024 & $518.70$ & $217.62^{+88.19}_{-91.86}$ & $-61.98 \pm 8.67$ & $-130.82 \pm 27.71$ & $-1.12^{+0.48}_{-0.32}$ & $<0.293$ & -- & -- & 18.7 \\
\hline
20231120A & 0.0368 & $438.90$ & $322.28^{+43.33}_{-39.88}$ & -- & -- & -- & $21.904^{+2.010}_{-2.523}$ & -- & -- & 37.2 \\
\hline
20231123B & 0.2625 & $396.70$ & $106.51^{+63.07}_{-65.40}$ & $379.37 \pm 18.10$ & $603.99 \pm 32.87$ & $7.02^{+2.49}_{-3.72}$ & $<1.666$ & -- & -- & 38.3 \\
\hline
20231220A & 0.3355 & $491.20$ & $185.52^{+87.89}_{-92.25}$ & $37.78 \pm 1.91$ & $89.31 \pm 9.39$ & $1.07^{+0.03}_{-0.77}$ & $0.641^{+0.073}_{-0.077}$ & -- & -- & 31.8 \\
\hline
20240104A & 1.3300 & $1351.00$ & $316.43^{+233.28}_{-230.12}$ & $27.45 \pm 9.08$ & $110.53 \pm 148.43$ & $1.13^{+0.61}_{-1.40}$ & $1.528^{+0.241}_{-0.253}$ & -- & -- & 11.3 \\
\hline
20240119A & 0.3760 & $483.10$ & $155.79^{+86.06}_{-89.85}$ & -- & -- & -- & $<3.132$ & -- & -- & 42.2 \\
\hline
20240123A & 0.9680 & $1462.00$ & $900.75^{+301.96}_{-308.60}$ & -- & -- & -- & $<7.586$ & -- & -- & 16.1 \\
\hline
20240203D & 0.0740 & $272.60$ & $113.20^{+42.58}_{-43.17}$ & $222.41 \pm 12.34$ & $271.76 \pm 22.63$ & $3.76^{+0.91}_{-1.69}$ & $0.653^{+0.084}_{-0.080}$ & $1542 \pm 56$ & $\lesssim 68.3$ & 18.4 \\
\hline
20240213A & 0.1185 & $357.40$ & $209.52^{+50.42}_{-49.44}$ & $-233.91 \pm 69.61$ & $-267.92 \pm 87.81$ & $-1.79^{+0.71}_{-0.69}$ & $6.533^{+0.773}_{-0.614}$ & -- & -- & 41.1 \\
\hline
20240215A & 0.2100 & $549.50$ & $352.81^{+75.54}_{-72.95}$ & $374.32 \pm 65.21$ & $527.60 \pm 97.05$ & $2.07^{+0.28}_{-0.72}$ & $6.775^{+0.987}_{-1.202}$ & -- & -- & 30.3 \\
\hline
20240224A & 0.3726 & $881.20$ & $611.08^{+120.73}_{-114.47}$ & $-114.50 \pm 71.21$ & $-137.60 \pm 136.15$ & $-0.32^{+0.14}_{-0.46}$ & $<0.178$ & -- & -- & 17.5 \\
\hline
20240229A & 0.2870 & $491.15$ & $234.82^{+85.29}_{-88.10}$ & $-6.15 \pm 0.06$ & $10.16 \pm 8.32$ & $0.16^{-0.01}_{-0.41}$ & $<0.196$ & -- & -- & 44.6 \\
\enddata
\tablecomments{All DMs are in $\mathrm{pc\,cm^{-3}}$ and all RMs in $\mathrm{rad\,m^{-2}}$. DM$_{\mathrm{host}}$ and RM$_{\mathrm{host}}$ have been scaled to the host-galaxy rest frame. Scattering timescales $\tau$ are in the observing frame, and scaled to 1 GHz assuming a scattering index $\alpha=4$. The $\tau$ reported for FRB 20221219A, which exhibits extreme scattering from intervening halos, was measured following \citet{faber2024}. FRB 20231120A and FRB 20231220A have intervening galaxies at impact parameters $b=34\,\mathrm{kpc}$ and $b=78\,\mathrm{kpc}$, respectively. The asterisk ($*$) marks FRB 20230814B, the only repeating source in this sample.}
\end{deluxetable*}

\begin{figure*}[]
  \centering
  \includegraphics[width=\textwidth]{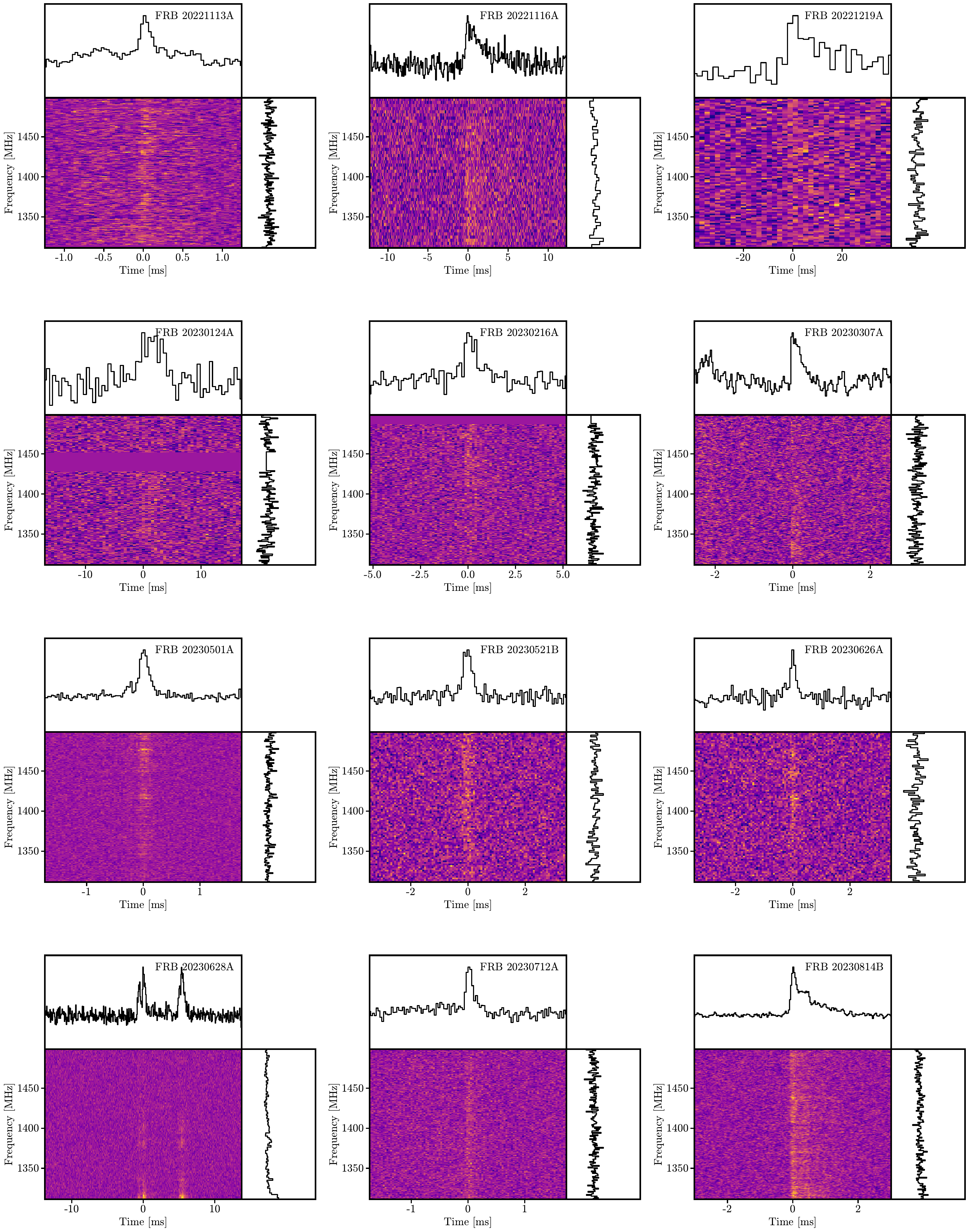}
   \caption{Summary plots for new DSA-110 FRBs. For each FRB, we show the dedispersed dynamic spectrum (center), the frequency-averaged time series (top), and the time-averaged spectrum (right). Data has been downsampled and normalized for visual clarity.}
    \label{catalog_1}
\end{figure*}

\begin{figure*}[t!]
  \centering
  \includegraphics[width=\textwidth]{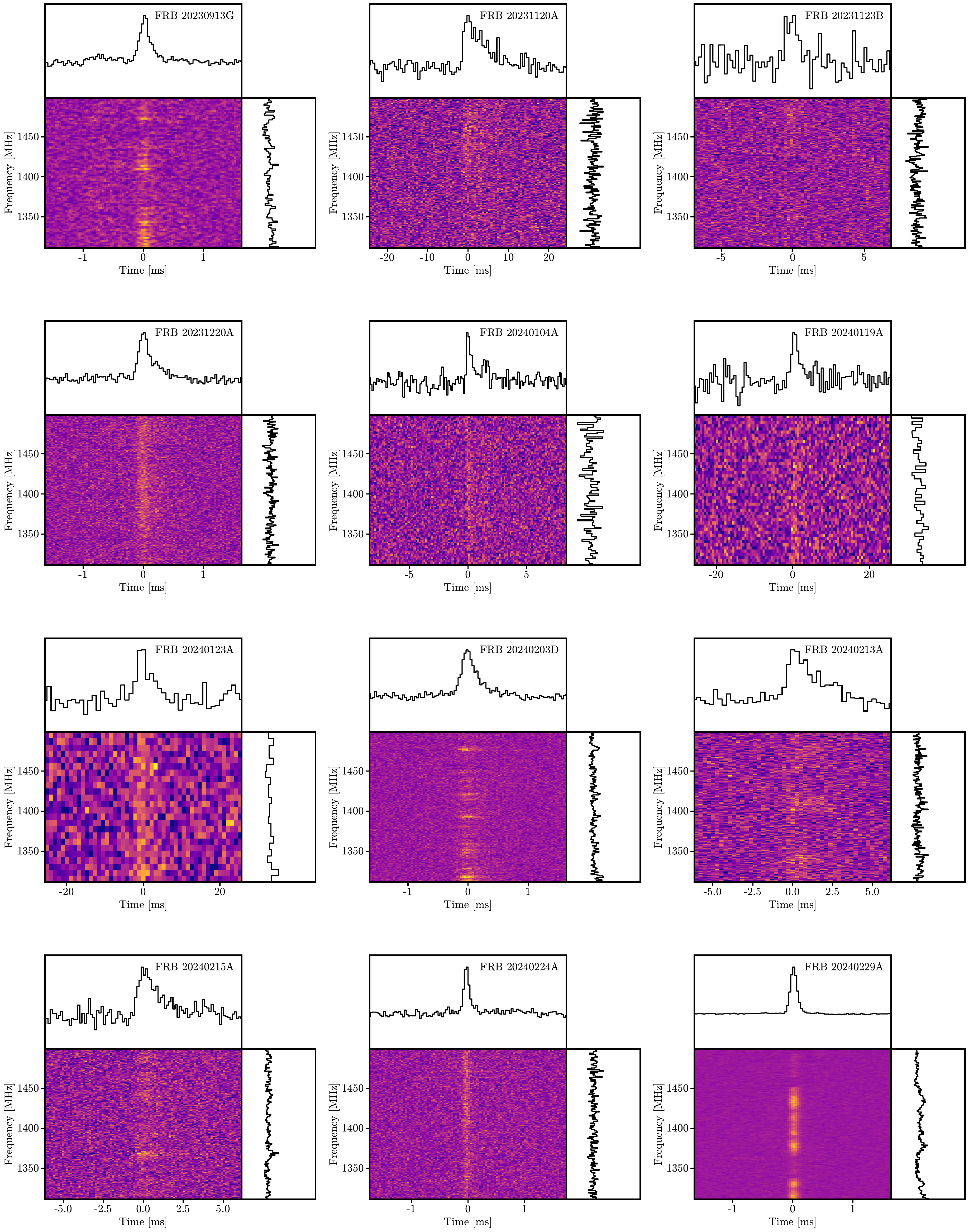}
    \caption{Same as Figure~\ref{catalog_1}.}
\end{figure*}

\begin{figure*}[]
  \centering
    \includegraphics[width=\textwidth]{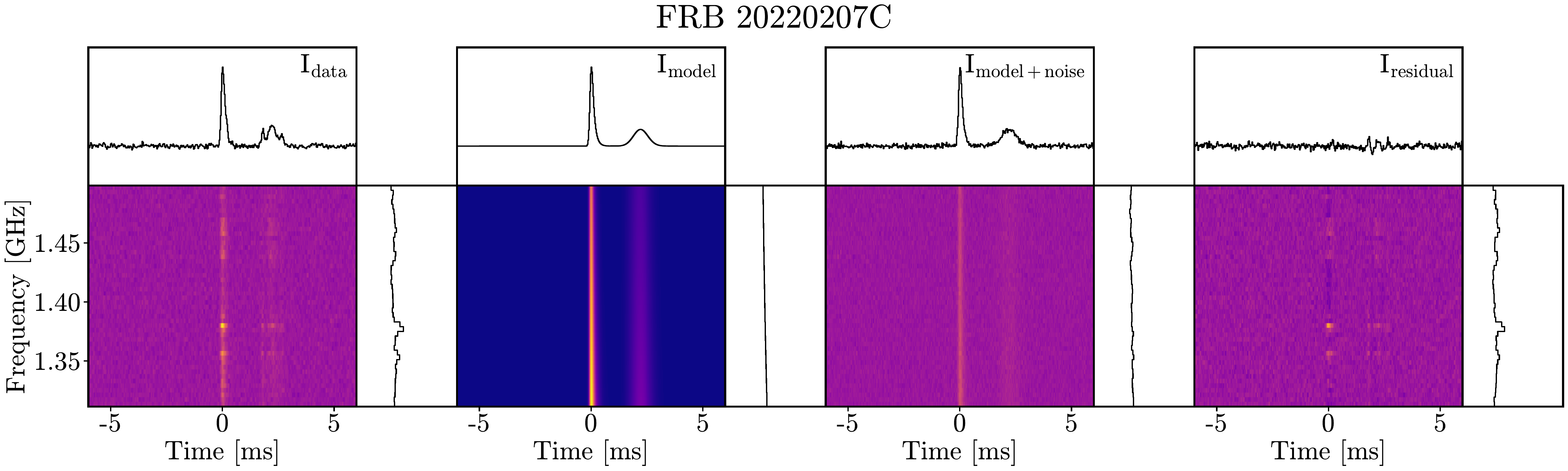}
    \includegraphics[width=\textwidth]{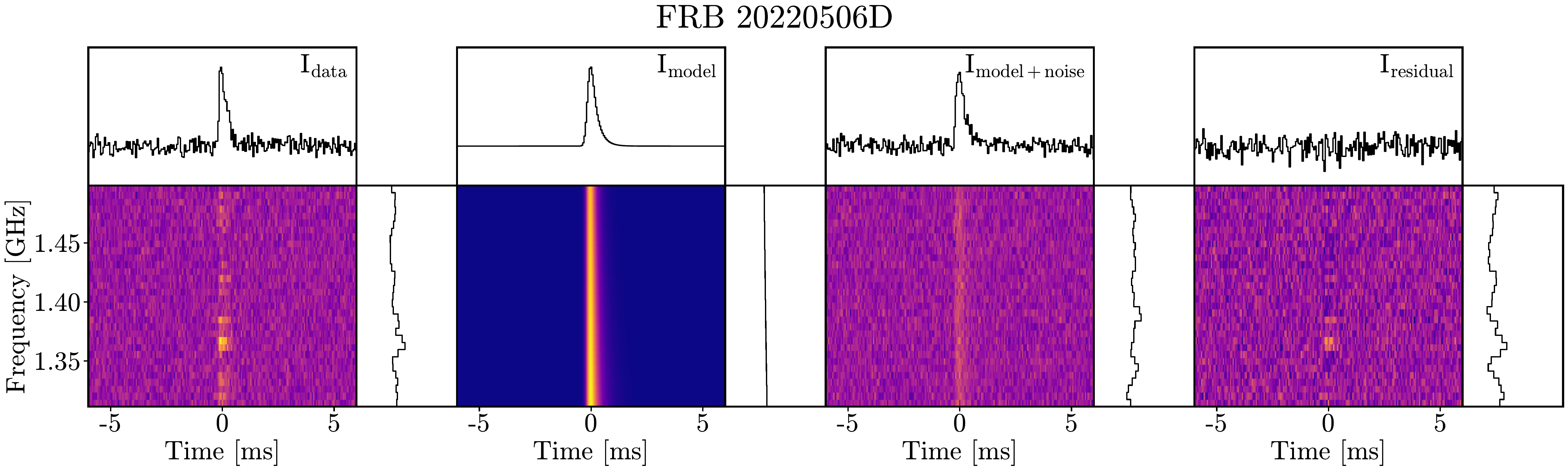}
    \includegraphics[width=\textwidth]{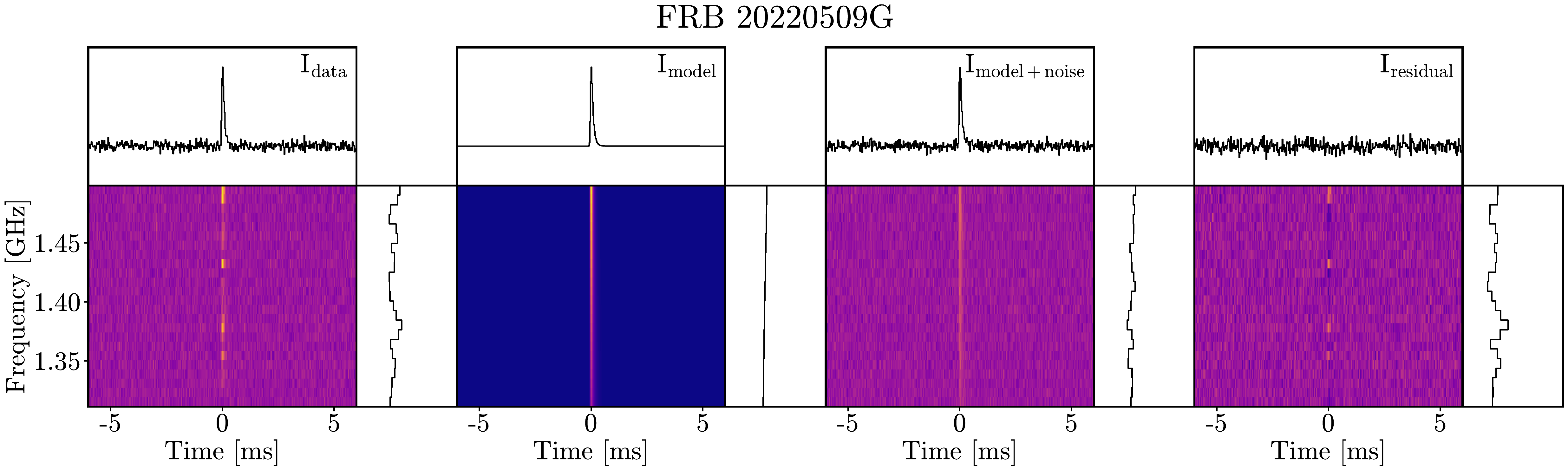}
    \includegraphics[width=\textwidth]{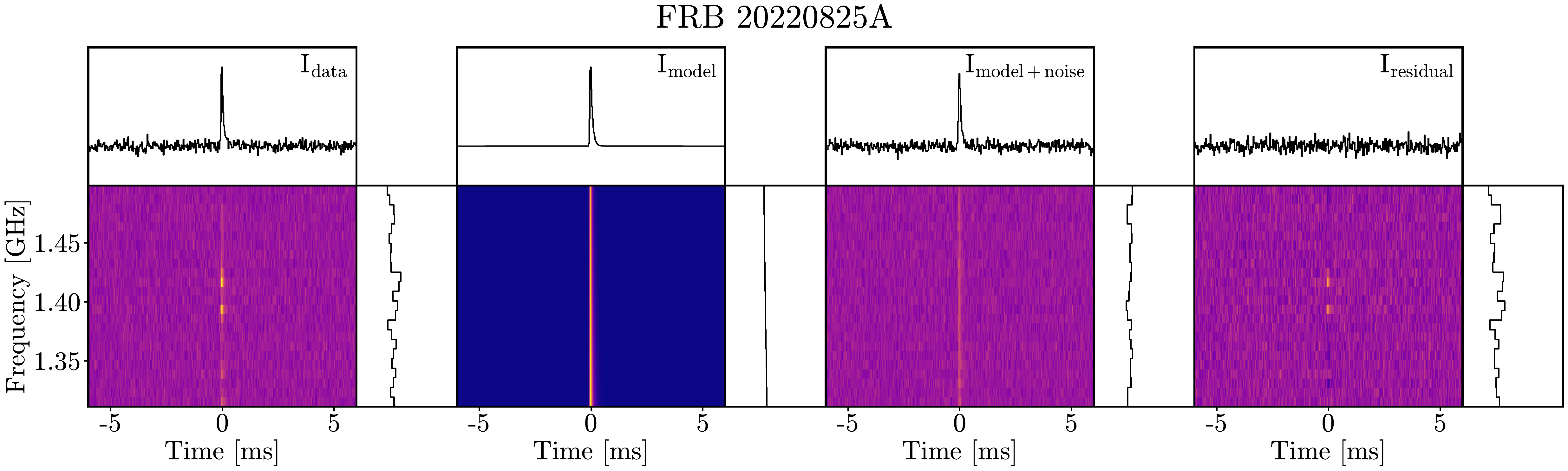}
    \caption{Scattering fits for DSA-110 FRBs \eqref{eq:scattering_model}. Column 1 shows the dynamic spectrum and timeseries of the observed FRB; Column 2 shows the best-fit model; Column 3 shows the same model with noise added at the level of the observed FRB; and Column 4 shows the residual $S_\nu(t) - D_\nu(t)$ between the model and the data.}
    \label{scatter_catalog_1}
\end{figure*}

\begin{figure*}[]
  \centering
    \includegraphics[width=\textwidth]{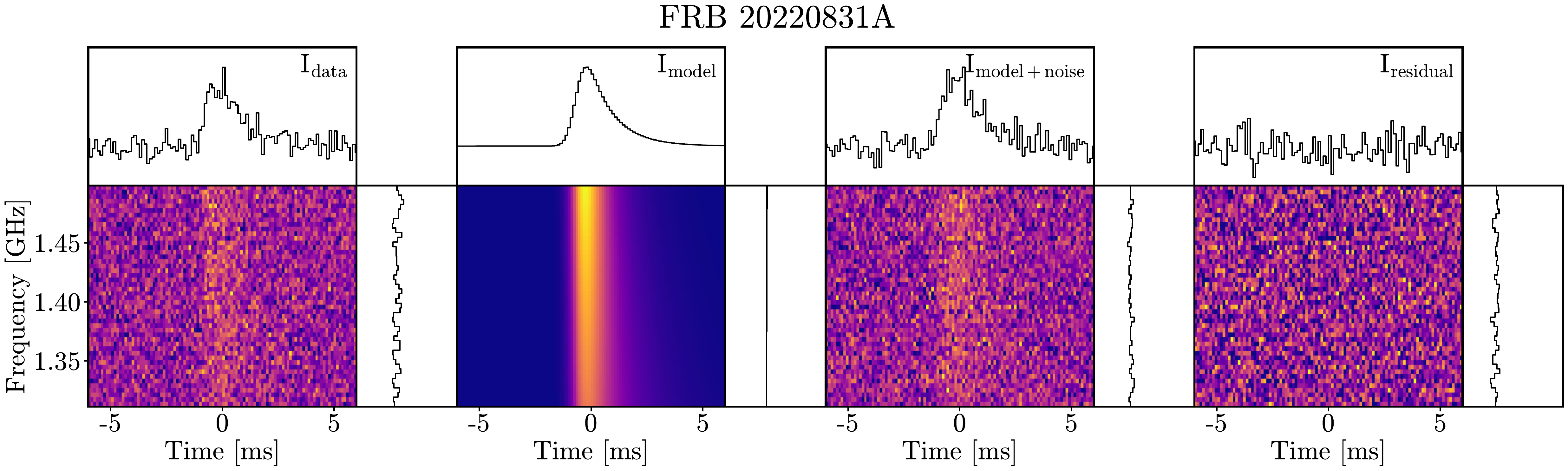}\\
    \includegraphics[width=\textwidth]{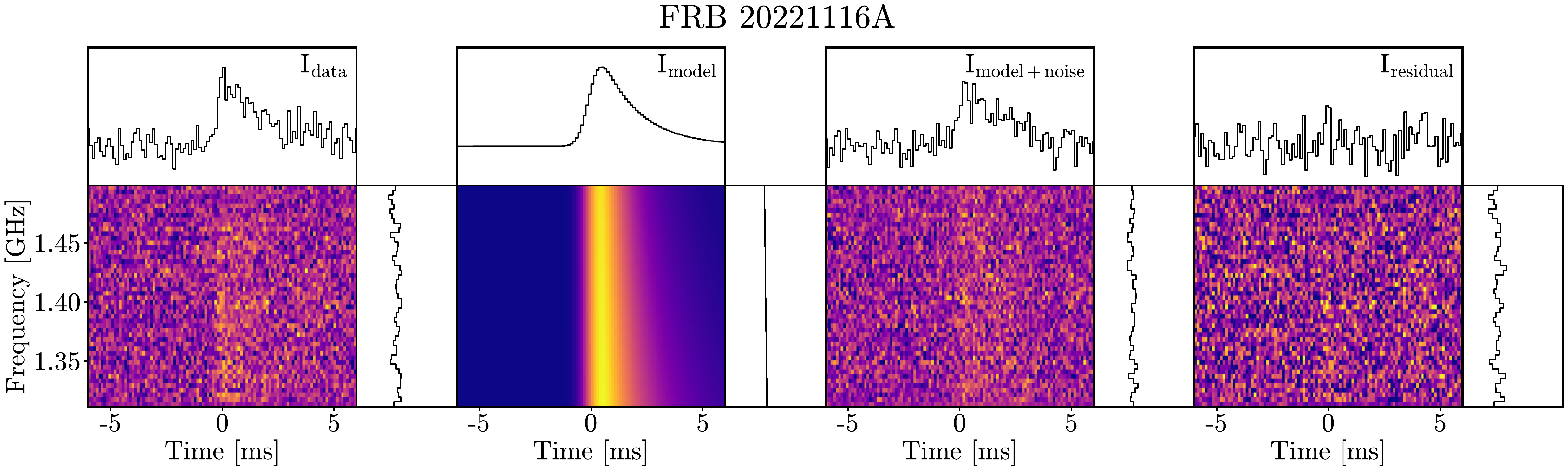}\\
    \includegraphics[width=\textwidth]{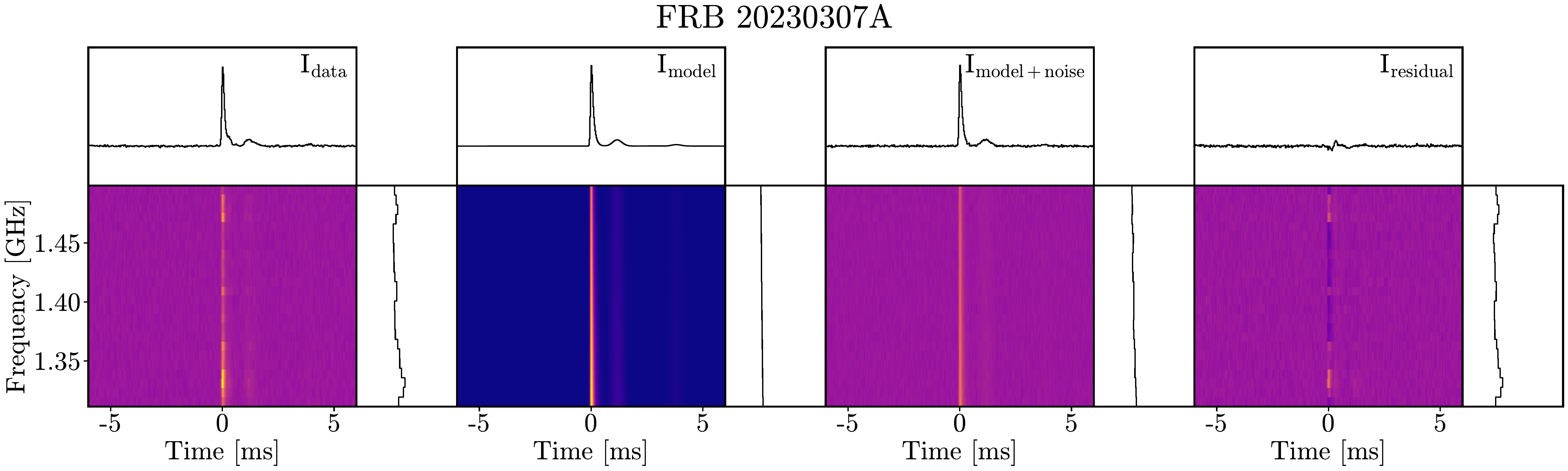}\\
    \includegraphics[width=\textwidth]{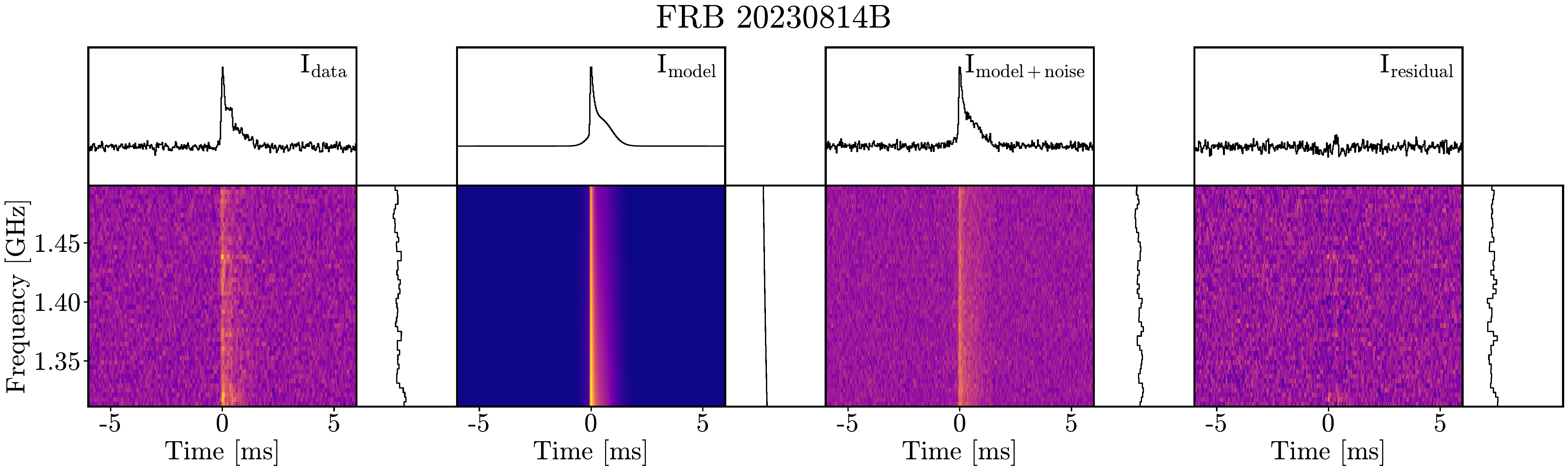}
    \caption{Same as Figure~\ref{scatter_catalog_1}.}
\end{figure*}

\begin{figure*}[]
  \centering
    \includegraphics[width=\textwidth]{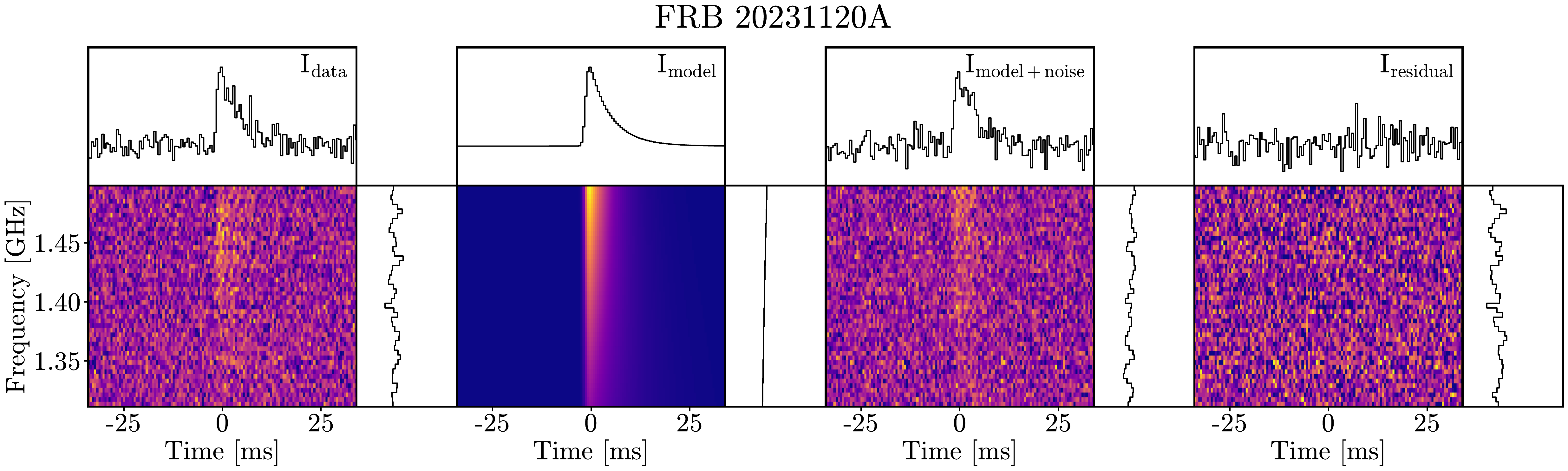}\\
    \includegraphics[width=\textwidth]{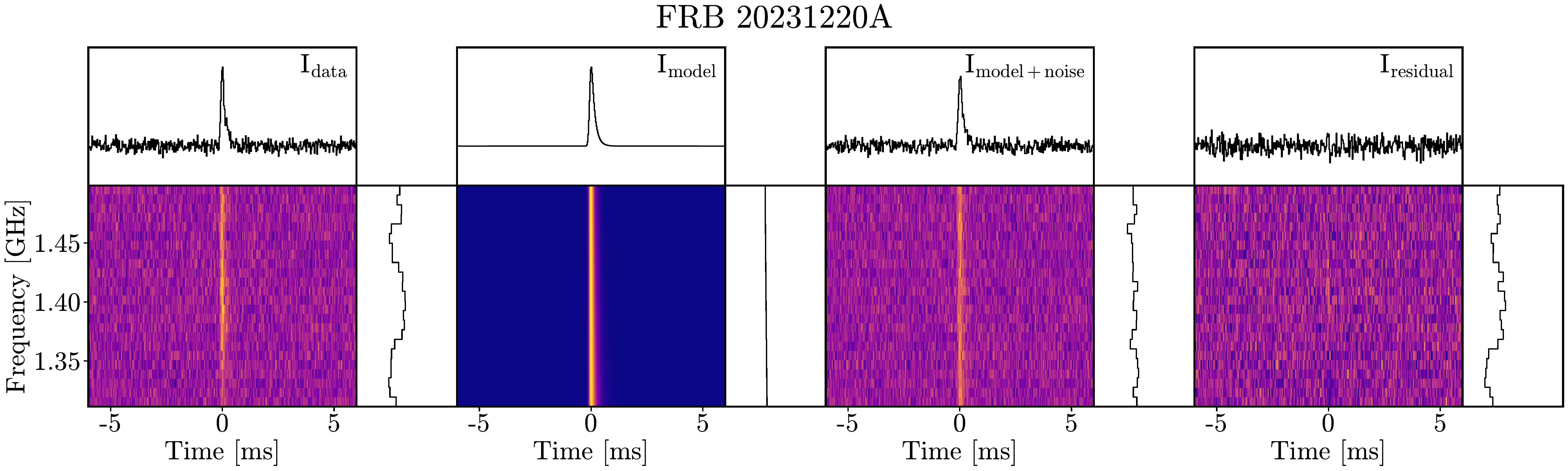}\\
    \includegraphics[width=\textwidth]{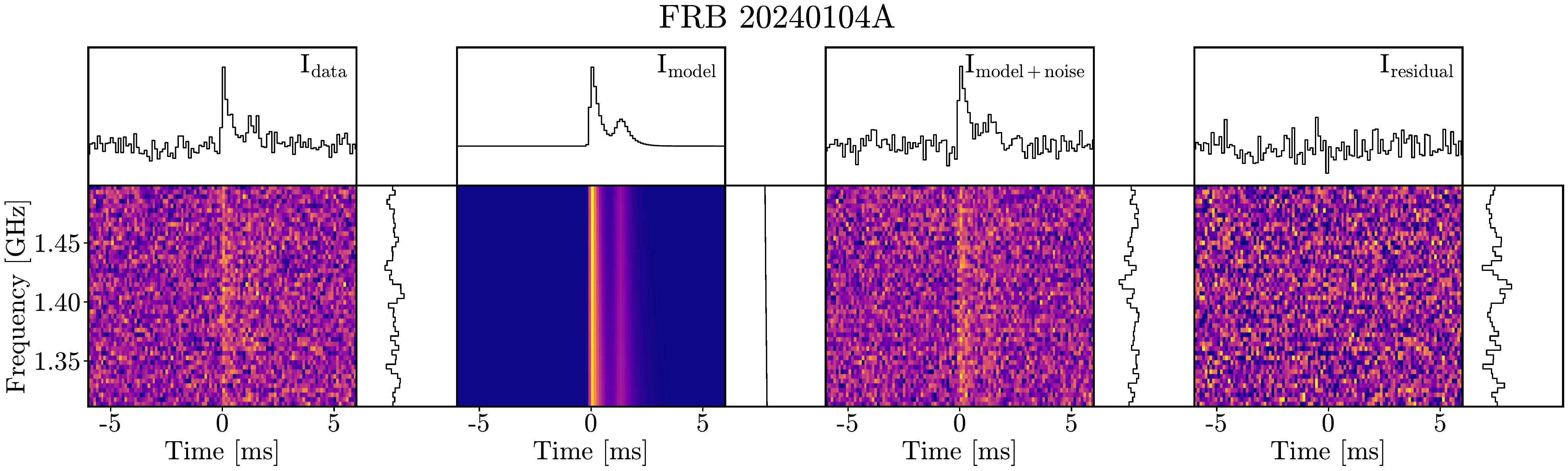}\\
    \includegraphics[width=\textwidth]{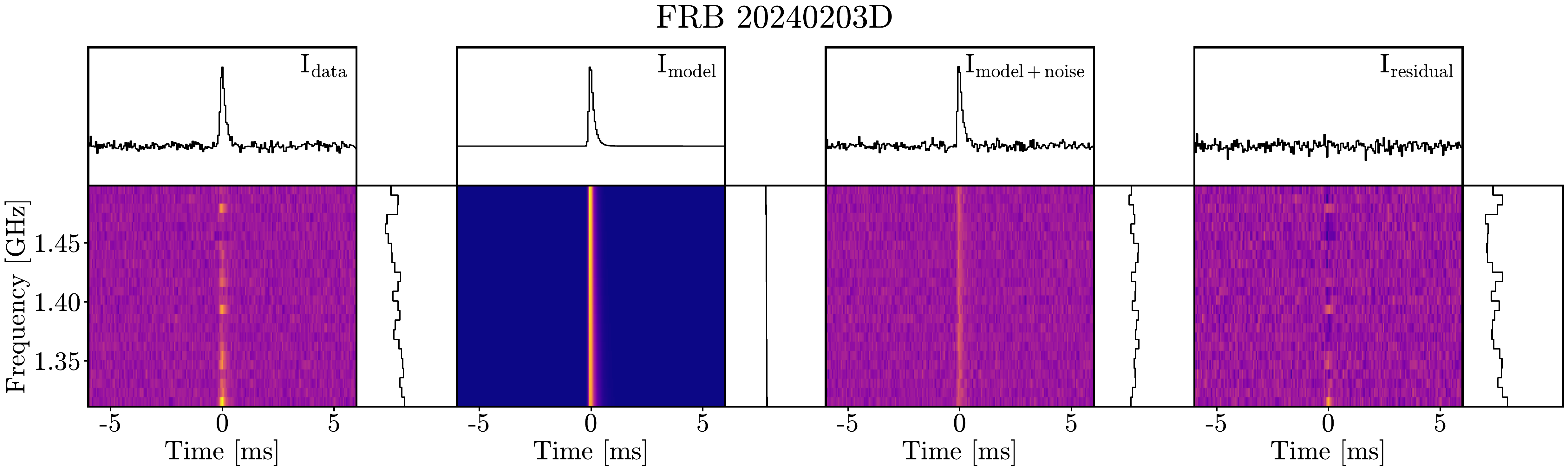}
    \caption{Same as Figure~\ref{scatter_catalog_1}.}
\end{figure*}

\begin{figure*}[]
  \centering
    \includegraphics[width=\textwidth]{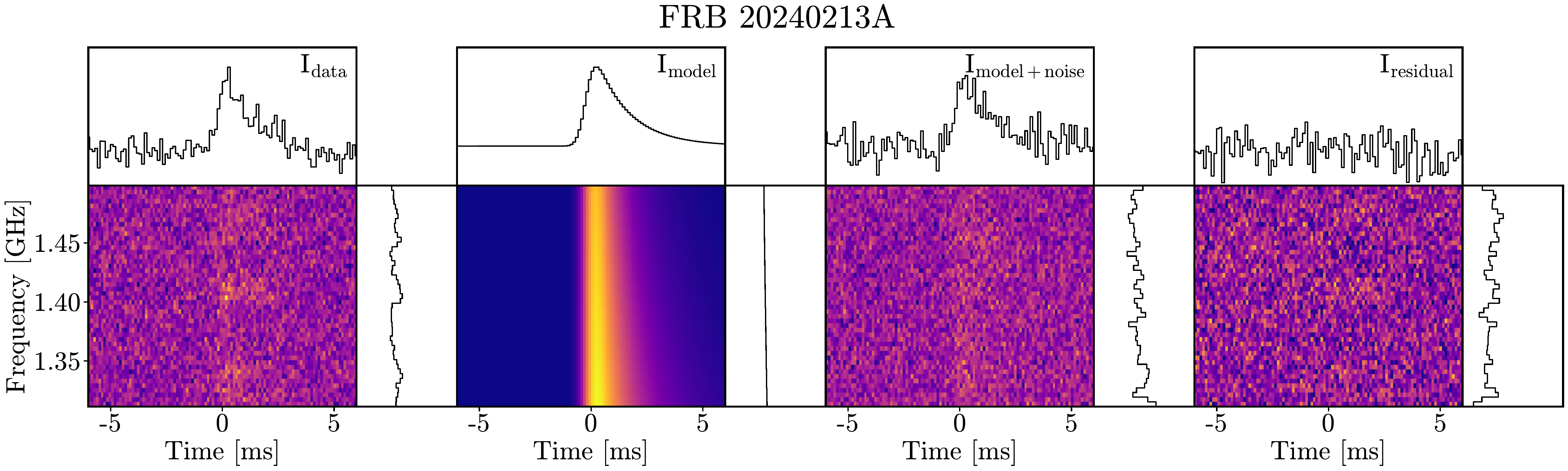}\\
    \includegraphics[width=\textwidth]{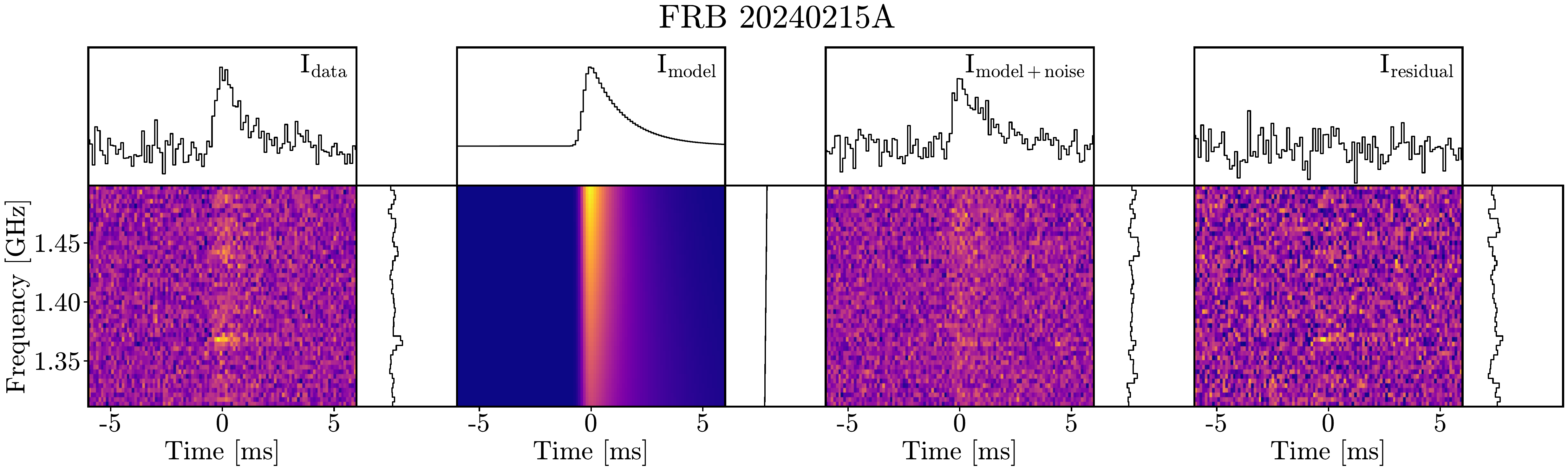}
    \caption{Same as Figure~\ref{scatter_catalog_1}.}
\end{figure*}

\begin{figure*}[]
    \centering
    \begin{minipage}{0.31\textwidth}
        \centering
        \includegraphics[width=\textwidth]{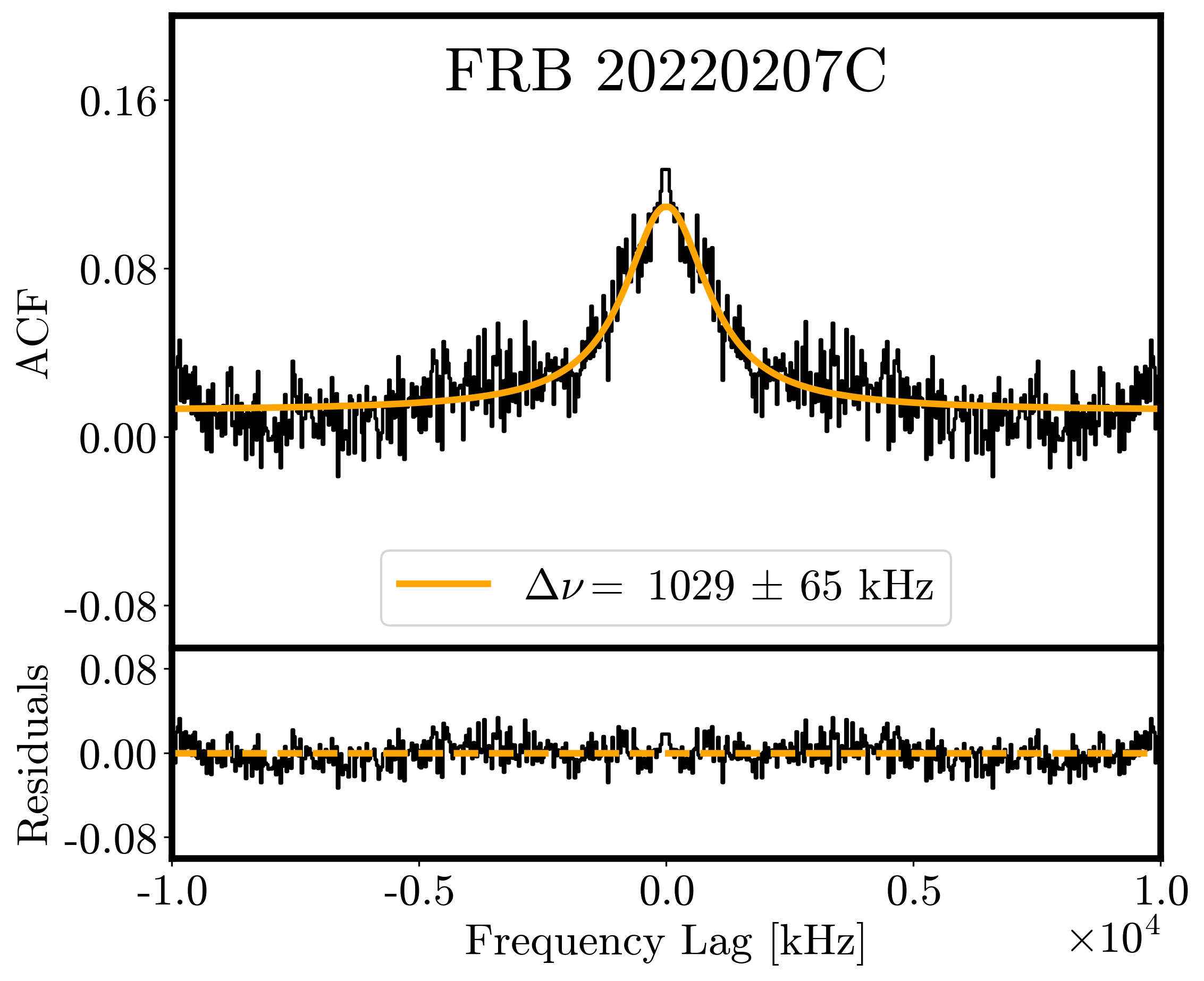}
    \end{minipage}
    \hfill
    \begin{minipage}{0.31\textwidth}
        \centering
        \includegraphics[width=\textwidth]{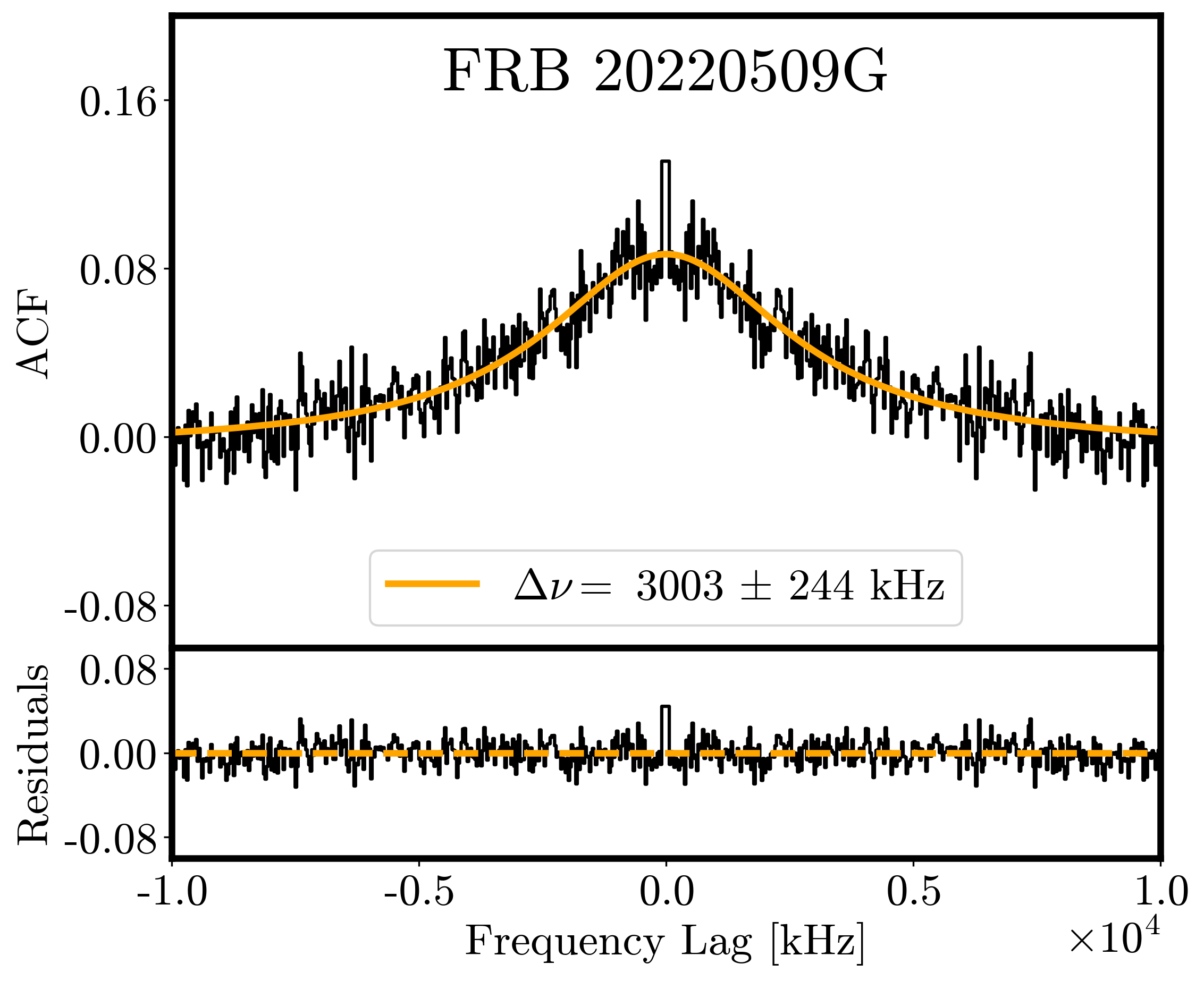}
    \end{minipage}
    \hfill
    \begin{minipage}{0.31\textwidth}
        \centering
        \includegraphics[width=\textwidth]{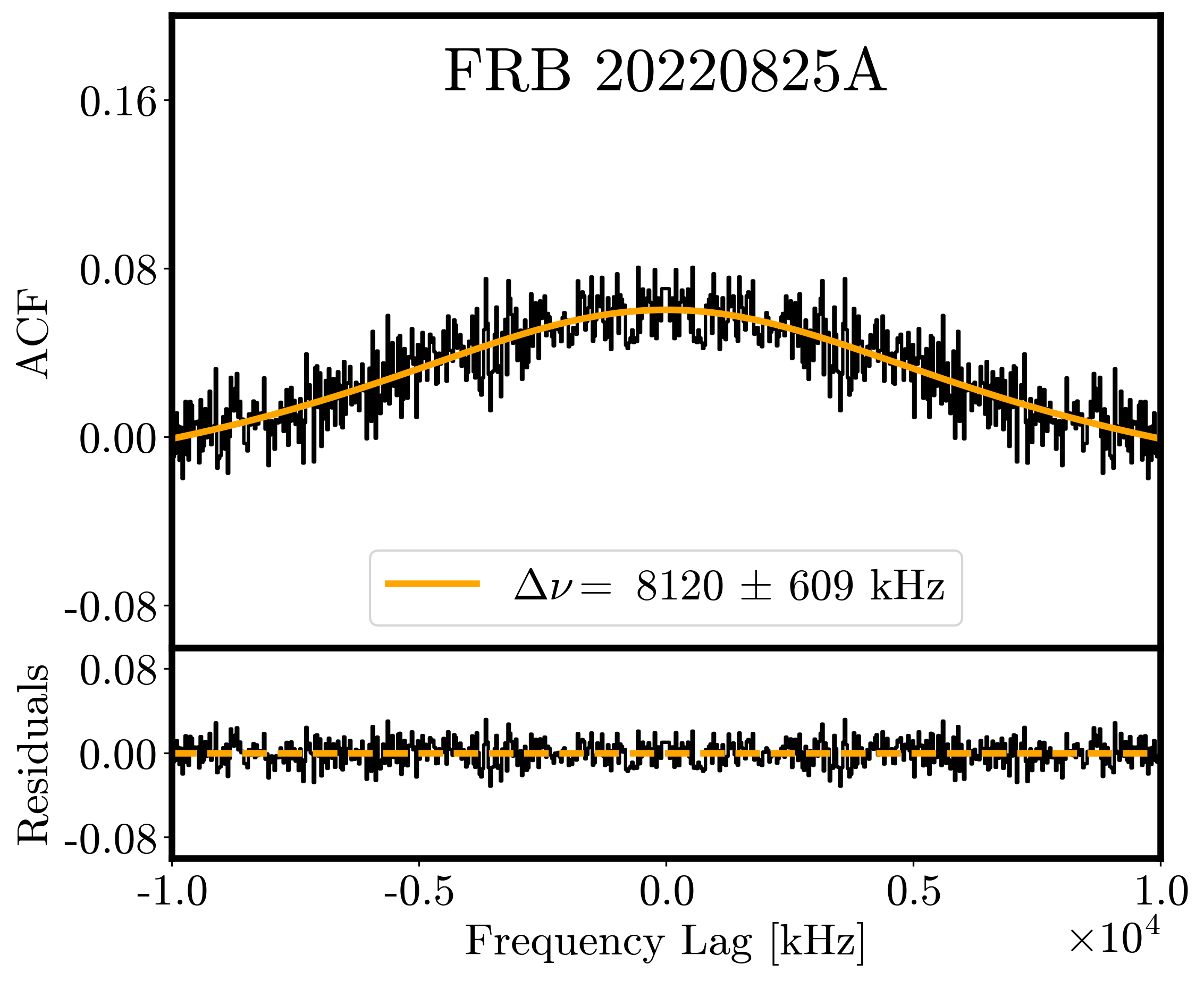}
    \end{minipage}

    \vspace{1em}

    % Row 2
    \begin{center}
        \begin{minipage}{0.31\textwidth}
            \centering
            \includegraphics[width=\textwidth]{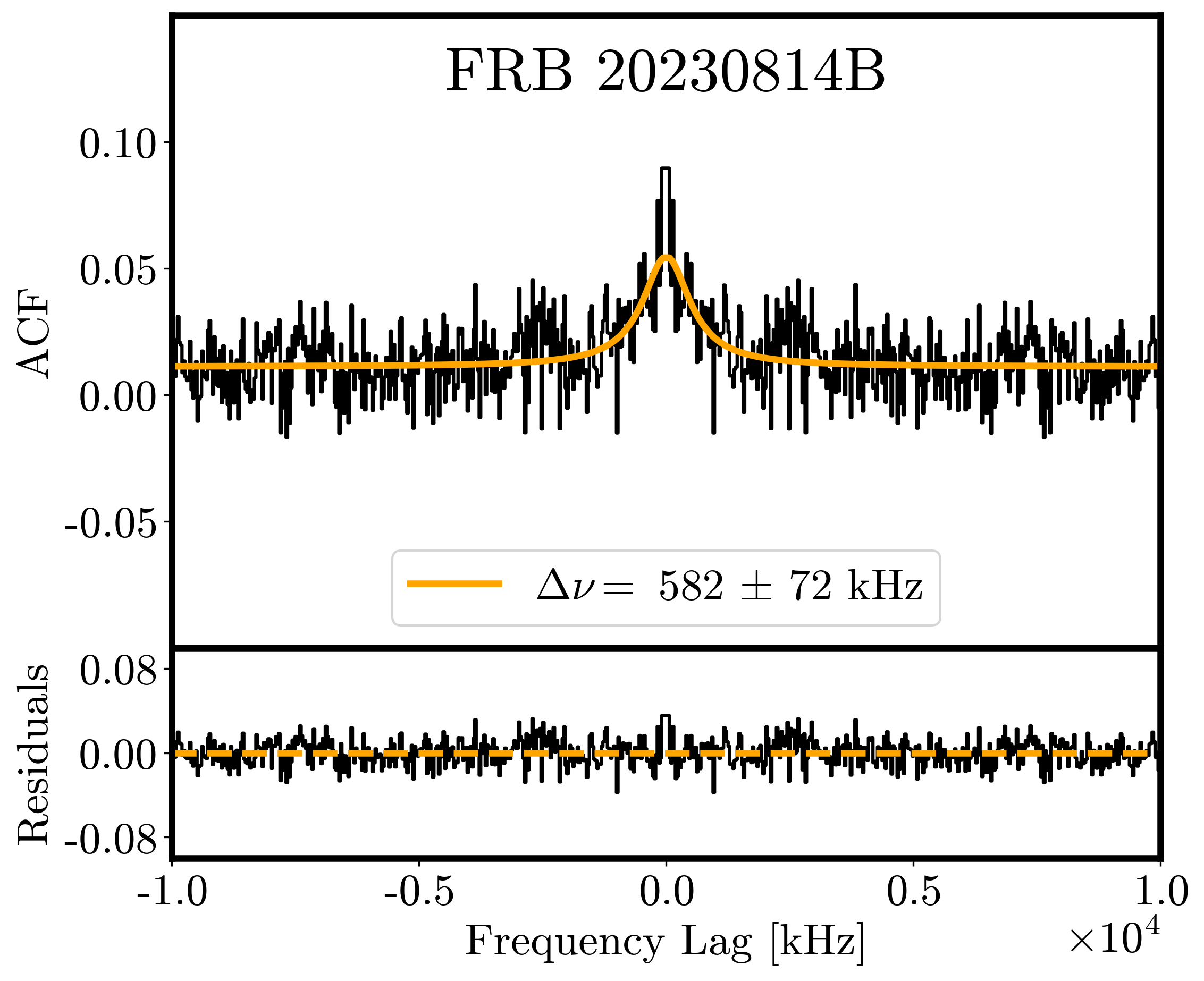}
        \end{minipage}
        \hspace{0.02\textwidth}
        \begin{minipage}{0.31\textwidth}
            \centering
            \includegraphics[width=\textwidth]{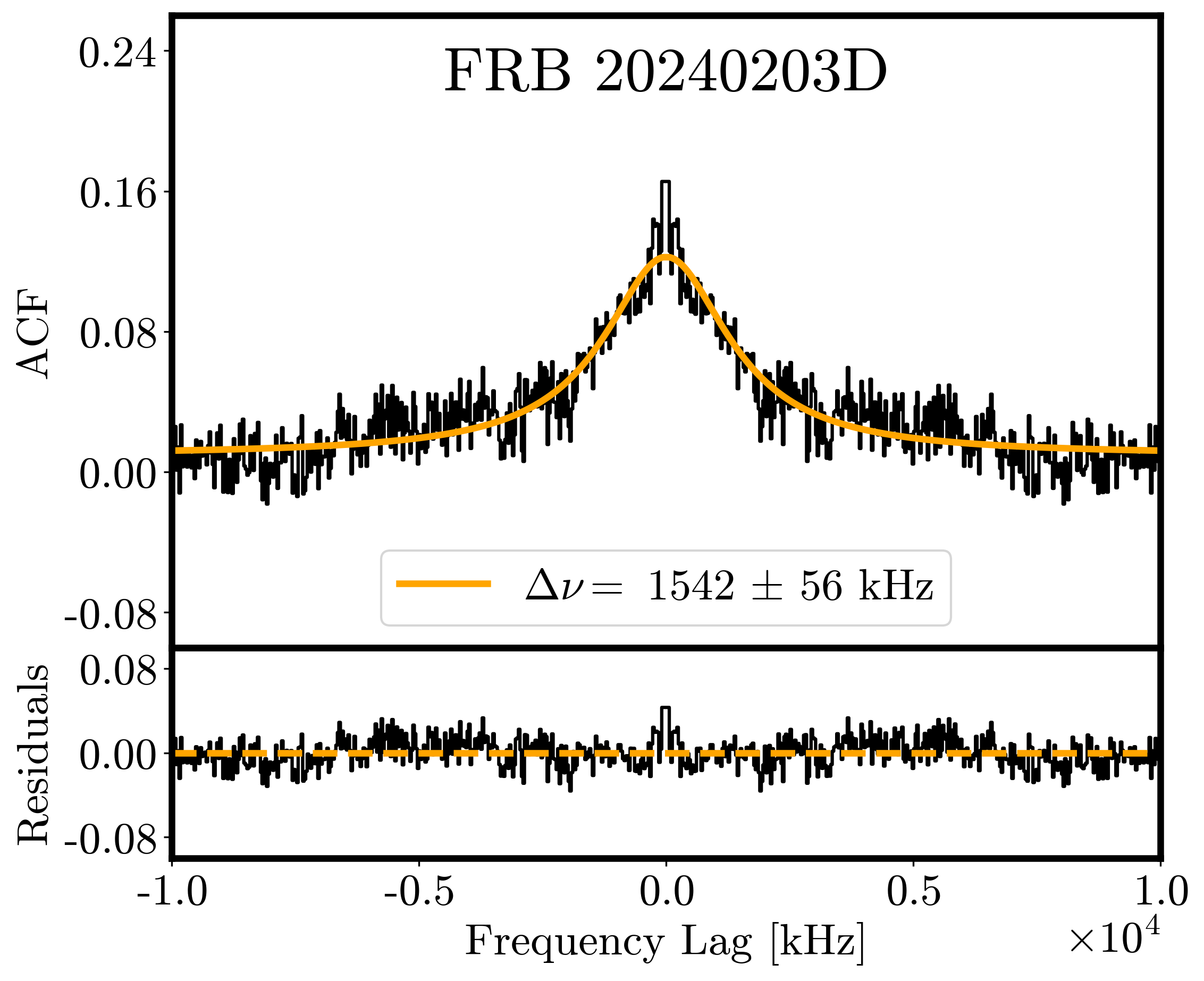}
        \end{minipage}
    \end{center}

    \caption{Scintillation measurements for DSA-110 FRBs with measurable pulse broadening. Each panel shows the intensity autocorrelation function (ACF) of the on-burst spectrum, with the zero-lag noise peak removed. For each FRB, a Lorentzian is fit to the central peak of the ACF using a least-squares fit, shown in orange.}
    \label{fig:appendix_scintillation}
\end{figure*}

\end{document}